\documentclass[]{emulateapj}

\usepackage{graphicx}
\usepackage{aas_macros}
\usepackage{amsfonts,amssymb}

\shorttitle{AHF: AMIGA's Halo Finder}
\shortauthors{Knollmann and Knebe}

\newcommand{\FOF}{\textsc{Fof}}
\newcommand{\SO}{\textsc{So}}
\newcommand{\DENMAX}{\textsc{Denmax}}
\newcommand{\SKID}{\textsc{Skid}}
\newcommand{\HOP}{\textsc{Hop}}
\newcommand{\BDM}{\textsc{Bdm}}
\newcommand{\VOBOZ}{\textsc{Voboz}}
\newcommand{\SUBFIND}{\textsc{Subfind}}
\newcommand{\MHF}{\textsc{Mhf}}
\newcommand{\AHF}{\textsc{Ahf}}
\newcommand{\MLAPM}{\textsc{Mlapm}}
\newcommand{\AMIGA}{\textsc{Amiga}}
\newcommand{\GADGET}{\textsc{Gadget}}
\newcommand{\MPI}{\textsc{MPI}}

\newcommand{\DomGrid}{\textsf{DomGrid}}
\newcommand{\DomRef}{\textsf{DomRef}}
\newcommand{\RefRef}{\textsf{RefRef}}
\newcommand{\LB}{\textsf{LB}}
\newcommand{\CPU}{\textsf{CPU}}

\newcommand{\dif}{\ensuremath{\mathrm{d}}}

\begin{document}

\title{AHF: AMIGA's Halo Finder}
\author{Steffen R. Knollmann\altaffilmark{1,2},
        Alexander Knebe\altaffilmark{1,2}}
\altaffiltext{1}{Departamento de F\'isica Te\'orica,
                 Modulo C-XI,
                 Facultad de Ciencias,
                 Universidad Aut\'onoma de Madrid,
                 28049 Cantoblanco, Madrid,
                 Spain}

\altaffiltext{2}{Astrophysikalisches Institut Potsdam,
                 An der Sternwarte 16,
                 14482 Potsdam,
                 Germany}
\email{steffen.knollmann@uam.es}

\begin{abstract} 

  Cosmological simulations are the key tool for investigating
  the different processes involved in the formation of the universe from
  small initial density perturbations to galaxies and clusters of galaxies
  observed today.  The identification and analysis of bound objects,
  halos, is one of the most important steps in drawing useful physical
  information from simulations.  In the advent of larger and larger
  simulations, a reliable and parallel halo finder, able to cope with the
  ever-increasing data files, is a must.  In this work we present the freely
  available \MPI\ parallel halo finder \AHF.  We provide a description of the
  algorithm and the strategy followed to handle large simulation data.
  We also describe the parameters a user may choose in order to
  influence the process of halo finding, as well as pointing out which
  parameters are crucial to ensure untainted results from the parallel
  approach.  Furthermore, we demonstrate the ability of \AHF\ to scale
  to high resolution simulations.

\end{abstract}


\section{Introduction}
\label{sec:introduction}

The identification and hierarchical grouping of `clumps' within large
sets of particles (may it be dark matter, star or gas particles)
produced by cosmological simulation codes is the objective in
halo-finding.  A variety of methods have been developed and have seen
many improvements over the years, however, at heart, all methods try to
identify peaks in the density field and group particles around those
peaks.

The classical method to identify isolated structures is the purely
geometrical `Friends-of-Friends' (\FOF) algorithm
\citep[e.g.][]{Davis1985} in which particles closer than a given scale
(the linking length) are connected.  The whole particle distribution
then separates into isolated regions where outside particles do not come
closer than the linking length.  A serious shortcoming of this method is
the danger of linking two blobs together via a `linking bridge'.
Additionally, with a fixed linking length it is impossible to identify
substructure within a \FOF\ halo. Many variants of this method have been
developed, trying to overcome the short-comings of the classical
algorithm, either by using adaptive linking lengths
\citep{Suginohara1992, vanKampen1995, Okamoto1999}, multiple linking
lengths \citep[][hierachical \FOF]{Klypin1999} or the inclusion of
earlier snapshot analyses \citep{Couchman1992, Summers1995,
Klypin1999}.

Most other halo finding methods employ an explicit measure of the
density field.  One of the first methods to do so is the Spherical
Overdensity (\SO) algorithm \citep{Press1974, Warren1992, Lacey1994}
which calculates the density from the distance to the $N$th nearest
neighbour.  It then searches for density peaks and grows spheres about
them until a certain overdensity threshold is reached iteratively
adapting the center to the center of the enclosed particles.

The \DENMAX\ algorithm \citep{Bertschinger1991, Gelb1994} uses a
grid to calculate the density field and then moves the particles on a path
given by the density gradient until a local density maxima is reached.
This artificially increases the separation of clumps and circumvents the
linking bridges, so that then a \FOF\ approach can be used to
collect halo particles.  Similar in spirit is the \HOP\ algorithm
\citep{Eisenstein1998} which employs a different way to move the
particles to a density maxima, avoiding the calculation of a density
gradient by instead `hopping' to a neighbouring particle associated with
the highest density.  The offspring of \DENMAX, \SKID\ \citep[see
e.g.][]{PhD_Stadel2001, Governato1997, Weinberg1997, Jang-Condell2001}
uses a Lagrangian density estimator similar to the one employed by the
\SO\ algorithm.

The \BDM\ method \citep{Klypin1997, Klypin1999} uses randomly
placed spheres with predefined radius which are iteratively moved to the
center of mass of the particles contained in them until the density
center is found.  This iteration process is also used in the \SO\
method.  Recently, the \VOBOZ\ \citep{Neyrinck2005} technique has been
described, which uses a Voronoi tessellation to calculate the local
density.  The \SUBFIND\ algorithm \citep{Springel2001a, Dolag2008} uses
in a first step a \FOF\ method and then looks for saddle points in the
density field within each \FOF\ group.

\citet{Gill2004} used the grid hierarchy generated by an adaptive mesh
refinement (AMR) code \citep[][\MLAPM]{Knebe2001} to construct a halo
finder, \MHF.  The grid hierarchy is built in such a way that the grid
is refined in high density regions and hence naturally traces density
contours.  This can be used to not only select halos, but also to
identify included substructure.  The AMR grid structure naturally
defines the halo-subhalo hierarchy on all levels, which is a mandatory
requirement for any state-of-the-art halo finder.

One further very important aspect in finding substructure is the pruning
of their associated particle lists to remove unbound particles to get a
dynamical description of the halo.  All current halo finding tools
(\SKID, \BDM, \SUBFIND, \VOBOZ, \MHF) perform this step, however the
degree to which the potential field is reconstructed to define the
binding energy varies.  Usually the halo is treated in isolation and
only recently methods have been proposed to handle the inclusion of
tidal effects to define the demarcation of subhalos
\citep[e.g.][]{Weller2005, Kim2006, Shaw2007}.

In this work we will describe the successor of \MHF\ named \AHF\ (the
\AMIGA\ Halo Finder).  \AMIGA\ aims to be the replacement for the cosmological
simulation code \MLAPM\ \citep{Knebe2001} and is capable of doing the
halo analysis during the course of the simulation.  However, \AHF\ can
also be used as a stand-alone halo finder and as such it is described in
this paper.  It features new reading routines which can handle large
simulation data in parallel and enhanced features, although the
principle algorithmic idea of \MHF\ is kept.

The paper is organized as follows.  \S\ref{sec:halo_finding} introduces
the method used to identify halos in cosmological simulations and
describes the parallel strategy.  In \S\ref{sec:testing} we apply \AHF\ 
and show the impact of the free parameters and verify that the parallel
approach does not introduce artefacts. We then compare \AHF\ to other
halo finders and to theoretical predictions in \S\ref{sec:comparison}.
\S\ref{sec:results} shows the scalability and stability of our results.
We conclude and summarize in \S\ref{sec:conclusions}.


\section{Halo finding the \AHF\ way}
\label{sec:halo_finding}

In this section we describe our algorithm for finding structures in
cosmological simulations.  \AHF\ is the successor of \MHF.  For a more
detailed description of the underlying principles of \MHF\ we refer the
reader to \citet{Gill2004} and \citet{PhD_Gill2005};  for reference,
especially on the configuration and compilation, the user manual
available on the \AHF\
website\footnote{\url{http://www.aip.de/People/AKnebe/AMIGA}} is also
helpful.  In this paper we focus on the parallel implementation and also
provide a study of the parameters of the algorithm.  However, we will
give a short description of the main ideas underlying \AHF\ in
\S\ref{subsec:ahf} and describe the parallelizing approach in
\S\ref{subsec:pahf}.  Finally we summarize the parameters in
\S\ref{subsec:AHFparameters}.

\subsection{\AHF}
\label{subsec:ahf}

In \AHF\ we start by covering the whole simulation box with a regular grid of a
user-supplied size (the domain grid, described by the \DomGrid\
parameter).  In each cell the particle density is calculated by means
of a triangular shaped cloud (TSC) weighing scheme
\citep{Book_Hockney1988}.  If the particle
density\footnote{Please note that the {\em particle} density is used,
not the mass density.} exceeds a given threshold (the refinement
criterion on the domain grid, \DomRef), the cell will be refined and
covered with a finer grid with half the cell size.  On the finer grid
(where it exists), the particle density is recalculated in every cell
and then each cell exceeding another given threshold (the refinement
criterion on refined grids, \RefRef) is refined again.  This is repeated
until a grid is reached on which no further cell needs refinement.

Note that the use of two different refinement criteria --- one for the
domain grid and one for the refinements --- has historical reasons routed
in the requirement that for a {\em cosmological simulation} the
interparticle separation of the initial particle distribution should be
resolved and hence  $\DomGrid = 2 N$, with $N^3$ being the total number
of particles.  However, this might lead to memory problems and hence the
ability to choose a coarser domain grid but to refine it more
aggressively was provided.  For {\em halo finding} the choice of the
domain grid is rather arbitrary.  We will test for the influence of
these parameters later on.

Following the procedure outlined above yields a grid hierarchy
constructed in such a way that it traces the density field and can then
be used to find structures in cosmological simulations:  Starting on the
finest grid, isolated regions are identified and marked as possible
halos.  On the next coarser grid again isolated refinements are
searched and marked, but now also the identified regions from the finer grid are
linked to their corresponding volume in the coarser grid.  Note that by
construction the volume covered by a fine grid is a subset of the
volume covered by the coarser grids.

\begin{figure}
 \includegraphics[width=84mm]{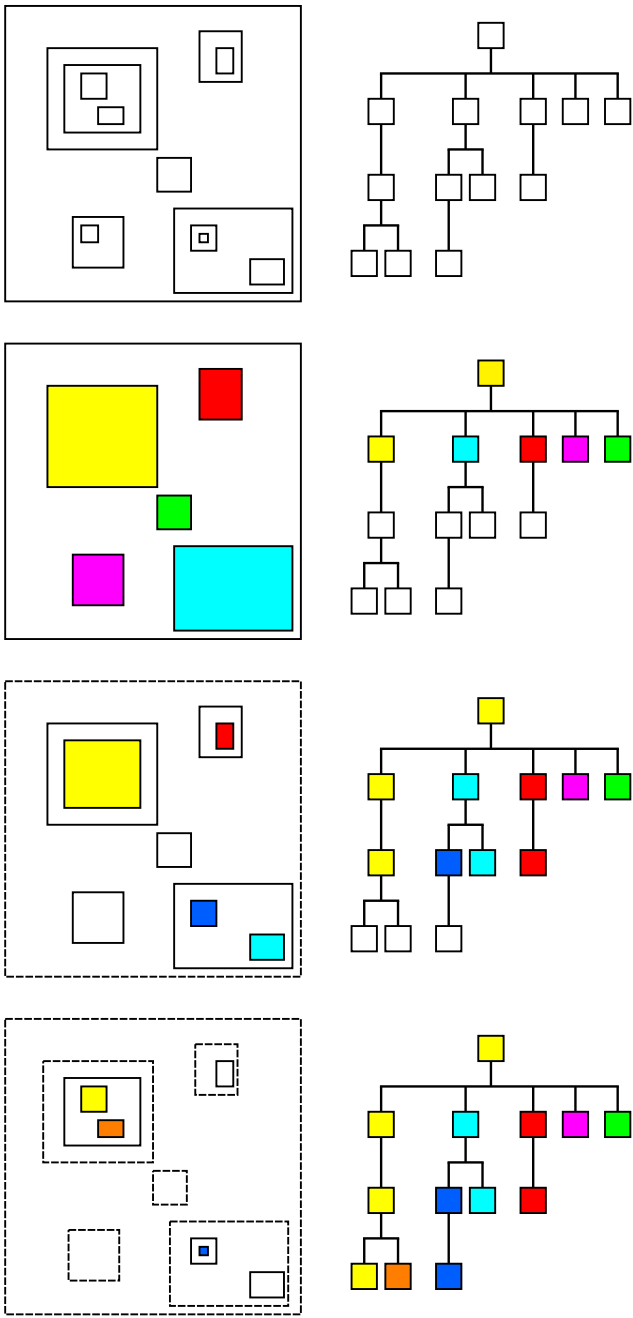}
  \caption{Here we illustrate the process of classifying a grid tree
into substructures.  In the top row, an arbitrary sample grid structure
is shown on the left and the corresponding grid tree on the right.  The
classification starts by labeling the coarsest grid as part of the host
and then proceeds to the next level.  This is depicted in the second row:
Five isolated grids are embedded and the one containing the most
particles is marked as the `host'.  The other four grids are then marked
as subhalos. This process is repeated for the next levels, always
deciding for each isolated grid whether is it the start of a new
substructure or part of the parent structure.}
  \label{fig:substructure}
\end{figure}

By following this procedure, a tree of nested grids is constructed
and we follow the halos from 'inside-out', stepping in density contour
level from very high densities to the background density.  Of course it
can happen that two patches which are isolated on one level link into
the same patch on the next coarser grid in which case the two branches
of the grid tree join.  The situation after this step is depicted in the
upper row of figure~\ref{fig:substructure}.

In a later step, once the grid forest is constructed, the classification
of substructure can be made.  To do this, we process each tree starting
from the coarsest level downwards to the finer levels, the procedure is
also illustrated in figure~\ref{fig:substructure}.  Once the finer
level splits up into two or more isolated patches, a decision needs to
be made where the main branch continues.  This is done by counting the
particles contained within each of the isolated fine patches and we
choose the one containing the most as the main branch, whereas the others
are marked as substructures\footnote{There are also different criteria
available and described in detail in the user manual  to make this
decision.}.  Note that this procedure is recursive and also applies to
the detection of sub-sub-structure.

Assuming now that the leaf of each branch of the grid tree corresponds
to a halo, we start collecting particles by first assigning all
particles on the corresponding isolated grids to this center.  Once
two halos `merge' on a coarser level, we tentatively assign all
particles within a sphere given by half the distance to the host halo
(here, the notation of substructure generated by the construction of
the grid tree enters).  We then consider the halo in isolation and
iteratively remove unbound particles, the details of this process are
described in appendix~\ref{app:AHFunbinding}.  Particles not bound to
a subhalo will be considered for boundness to the host halo.  By doing
this for all prospective centers, we construct a list of halos with
their respective particles; note that subhalos are included in their
parent halo by default, but there is, however, an option to not
include them. 

The extent of the haloes is defined such that the virial radius is given
by
\begin{equation}
	\bar{\rho}(r_{\mathrm{vir}}) = \Delta_{\mathrm{vir}}(z) \rho_b
\end{equation}
where $\bar{\rho}(r)$ denotes the overdensity within $r$ and $\rho_b$ is
the background density.  Hence the mass of the halo becomes
\begin{equation}
	M_{\mathrm{vir}} = 4 \pi \rho_b \Delta_{\mathrm{vir}}(z)
	                   r_{\mathrm{vir}}^3/3 \quad .
\end{equation}
Note that the virial overdensity $\Delta_{\mathrm{vir}}$ depends on the
redshift and the given cosmology, it is however also possible to choose
a fixed $\Delta$.  This definition for the extent of a halo does not
necessarily hold for subhalos, as it can happen that the overdensity never
drops below the given threshold.  The subhalo is therefore truncated
at higher overdensities, given by a rise in the density profile.

We would like to note that host halos (or in other words field halos)
initially include all particles out to the first isodensity contour
that fulfills the criterion $\rho_{\rm iso} < \Delta_{\mathrm{vir}}(z)
\rho_b$. Then the same procedure as outlined above is applied, i.e.
the halo is considered in isolation and unbound particles are
iteratively removed.

The (bound) particle lists will then be used to calculate canonical
properties like the density profile, rotation curve, mass, spin, and
many more.  After the whole halo catalog with all properties has been
constructed, we produce three output files: The first one contains the
integral properties of each halo, the second holds for each halo
radially binned information and the third provides for each halo a
list of the IDs of all particles associated with this halo.

As this algorithm is based on particles, we natively support
multi-mass simulations (dark matter particles as well as stars) and
also SPH gas simulations.  For \AHF\ they all are `just' particles and
we can find halos based on their distribution, however, for the gas
particles of SPH simulations we also consider their thermal energy in
the unbinding procedure.  Even though, for instance, the stellar
component is much more compact than the DM part, this does not pose a
challenge to \AHF\ due to its AMR nature: the grid structure is
generated based on {\em particle} densities rather than
{\em matter} densities that are obviously dominated by dark matter
particles.

To summarize, the serial version of \AHF\ exhibits, in principle, only
few parameters influencing the halo finding.  The user has three
choices to make, namely the size of the domain grid (\DomGrid), the
refinement criterion on the domain grid (\DomRef) and the refinement
criterion on the refined grids (\RefRef) need to be specified. While
all these three parameters are mainly of a technical nature they
nevertheless influence the final halo catalogues obtained with
\AHF. The code utilizes isodensity contours to locate halo
centres as well as the outer edges of the objects. Therefore,
{\DomGrid} along with {\DomRef} sets the first such isodensity level
whereas {\RefRef} determines how finely the subsequent isodensity
contours will be spaced: if {\RefRef} is set too large an object may
not be picked up correctly as \AHF\ samples the density
contours too coarsely.  We refer the reader to a more indepth study of
the influence of these parameters on the results in
Section~\ref{sec:testing}.

\subsection{Parallel Approach}
\label{subsec:pahf}

We will now describe the approach taken to parallelize the halo-finding.
In \S\ref{subsubsec:domdecomp} we describe the way the volume is
decomposed into smaller chunks and we then elaborate in
\S\ref{subsubsec:lb} how the load-balancing is achieved before
ending in \S\ref{subsubsec:caveats} with some cautionary notes on the
applicability of this scheme.

\subsubsection{Volume Decomposition}
\label{subsubsec:domdecomp}

There are many ways to distribute the workload to multiple
processes.  In some halo finders this is done by first generating
particle groups with a \FOF\ finder which can then independently
processed \citep[e.g.][]{Kim2006}.  However, we chose to `blindly' split
the whole computational volume to multiple CPUs.  This is done because
we plan to implement \AHF\ as part of the simulation code \AMIGA\ and
not only as a stand-alone version\footnote{Note that for the serial
version this is already the case.}; the same as \MHF\ is also integrated into
\MLAPM\ and can perform the halo finding `on the fly' during the runtime
of a simulation.  For the simulation code, a proper volume decomposition
scheme is required and hence we are using the approach described here.
However, it should be noted that, in principle, \AHF\ is capable of
working on any user defined sub-set of a simulation box and as such is
not tied to the employed decomposition scheme.

\begin{figure}
 \includegraphics[width=84mm]{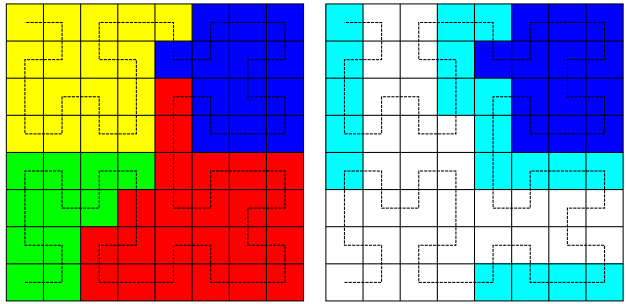}
  \caption{A sample volume decomposition is shown in the left panel, the
total volume is divided into four segments along a Hilbert curve (dashed
line) on the load-balance grid.  Here we choose $\LB = 3$ and only
show a two dimensional sketch for the sake of clarity.
The right panel shows for one selected segment, that is assigned to one
\MPI\ process, which additional boundary volume elements will be copied to the
task.}
  \label{fig:decomposition}
\end{figure}

The volume decomposition is done by using a space-filling curve (SFC),
which maps the three spatial coordinates $(x,y,z)$ into a
one-dimensional SFC-index $i$; an illustration is given in the left
panel of figure~\ref{fig:decomposition}.  As many other workers have done
before~\citep[e.g.][]{Springel2005, Prunet2008} we use the Hilbert-curve.
A new parameter, \LB, of \AHF\ regulates on which level this ordering is
done; we then use $2^{\LB}$ cells per dimension; it is important to
remark that the grid used for load-balancing is distinct from the domain
grid and therefore an additional parameter.  Each process will then
receive a segment, described by a consecutive range of indices
$i_\mathrm{start}\ldots i_\mathrm{stop}$, of
the SFC curve and hence a sub-volume of the whole computational box.
The Hilbert curve has the useful property to preserve locality and also
to minimize the surface area.  It is in this respect superior to a
simple slab decomposition, however at the cost of volume chunks with a
complicated shape.

In addition to its segment of the SFC curve, each process will also
receive copies of the particles in a buffer zone around its volume with
the thickness of one cell of the \LB-grid (cf.\ the right panel of
figure~\ref{fig:decomposition}). With these particles, each CPU can then
follow the standard recipe described above (\S\ref{subsec:ahf}) in
finding halos.  The thickness of the boundary is hence an important
quantity that is supposed to counteract the situation in which one halo
is situated close to the boundary of two processes.  As there is no
further communication between the tasks, for reasons we will allude to
below, the buffer zones need to be large enough to encompass all
relevant particles.  To fulfill this requirement, the thickness of the
boundary zone, given by the \LB\ parameter, should obey the relation
\begin{equation}
  2^{\LB} \lesssim \frac{B}{R_\mathrm{vir}^\mathrm{max}}
\label{eq:boundary}
\end{equation}
where $B$ is the size of the simulation box and
$R_\mathrm{vir}^\mathrm{max}$ is the radius of the largest objects of
interest.  Note that each process only keeps those halos whose centers
are located in the proper volume (as given by the SFC segment) of the
process.

\subsubsection{Load-balancing}
\label{subsubsec:lb}

To subdivide the SFC curve, different schemes can be employed, influencing
the quality of the achieved load-balancing; this is ultimately measured
by the difference of the run times of the different processes.  We chose
to use a scheme that distributes the particles evenly between the
processes, e.g.  each CPU will require the same amount of storage for
the particles.  This can of course lead to vastly different volumes
assigned to each task (cf.~figure~\ref{fig:decomposition}), but we will
show in \S\ref{subsec:res_lb} that this simple scheme can provide
reasonable results.

We therefore segment the SFC curve into chunks containing the same
(within the precision allowed by the coarseness of the load balance grid
given by \LB) number of particles.  As with this scheme (or any scheme
based on segmenting the volume along a SFC curve) it can happen that
objects are cut between different processes, the grid tree construction
would be very communication intensive.  To circumvent this, we use the
inclusion of a buffer zone alluded to above.

Note that after the duplication of boundary particles the balance of
particles might shift, which is especially prominent for very
inhomogeneous particle distributions, as found, for example, in
simulations with a small box size.  Since the decision of how to segment
the SFC curve is modular, it is very easy to implement a different scheme,
we plan on investigating into further criteria in future work.

\subsubsection{Caveats}
\label{subsubsec:caveats}

As we described above, we require that a single halo is processed
by one task, we do not yet allow for parallelism on a halo level.  Hence
we introduced the duplication of particles, which can lead to
an unfortunate distribution of the work-load:  Large halos may require more
time to be processed than many smaller halos.  To counteract this, we
provide the option of dividing the halo finding into a splitting
and an analysing step.  Only the splitting step needs to be performed by
a parallel task in which each process will dump its particles to a
restart file.  All those can then be analysed independently.

This becomes important for large simulations, containing hundreds of
million of particles.  On smaller simulations (up to $512^3$) we find
that the runtime unbalance is not very significant in absolute terms and
a direct parallel approach is very well feasible, we will discuss this
further in \S\ref{subsec:res_lb}.

\subsection{The parallel parameters of \AHF}
\label{subsec:AHFparameters}

To summarize, besides of the three already existing parameters
(cf. Section~\ref{subsec:ahf}) the user of the MPI-enabled version of
\AHF\ has to choose two additional parameters: the number of {\CPU}s
will influence the total runtime and is the parameter that can be
adapted to the available resources.  Very crucial, however, is the
size of the boundary zones, given by \LB, which has a significant
influence on the reliability of the derived results and must be chosen
carefully (cf.\ equation \ref{eq:boundary}).

All of the \AHF\ parameters are of a more technical nature and
hence should not influence the final halo catalogue. However, we like
to caution the reader that inappropriate choices can in fact lead to
unphysical results. The following Section~\ref{sec:testing} therefore
aims at quantifying these influences and deriving the best-choice
values for them.


\section{Testing}
\label{sec:testing}

In this section we are discussing the impact of the free parameters of
\AHF\ on the halo finding.  In \S\ref{subsec:simus} we first
introduce the ensemble of simulations we used and describe the tests.
To gauge the influence of numerical effects, we perform one additional
test in \S\ref{subsec:numerics}, before systematically varying the
parameters in \S\S\ref{subsec:domgrid}-\ref{subsec:test_parallel}.

\subsection{Performed tests}
\label{subsec:simus}

We use two different sets of simulations, summarized in
table~\ref{tab:simulations}.  To investigate the impact of
the free parameters of the algorithm and to produce comparative figures
of the serial version to the parallel version, we employ the simulations
containing $256^3$ particles contained in set~1. The boxes have different sizes, 
namely $20\,h^{-1}\,\mathrm{Mpc}$, $50\,h^{-1}\,\mathrm{Mpc}$ and
$1.5\,h^{-1}\,\mathrm{Gpc}$.  The first simulation belongs to the set of
simulations described in more detail in \citet{Knebe2008}. The
particulars of the second simulation are described elsewhere (Power et
al. in prep.)  and the last one is described in \citet{Wagner2008}.
Note that with this particle resolution we can still use the serial
version to produce halo catalogs.  An additional set of simulations is
later used (cf. \S\ref{subsec:scalability}) to investigate the scaling
behaviour of \AHF\ with the number of particles.  The three simulations
of the $936\,h^{-1}\,\mathrm{Mpc}$ box forming set~2 were provided to us
by Christian Wagner.

\begin{deluxetable}{llcc}
  \tablecaption{Summary of the simulation parameters. \label{tab:simulations}}
  \tablehead{
     \colhead{} &
     \colhead{Name} &
     \colhead{$B [h^{-1}\,\mathrm{Mpc}]$} & 
     \colhead{$N_\mathrm{part}$}
  }
  \startdata
  Set 1 & B20 & $20$ & $256^3$ \\
        & B50 & $50$ & $256^3$ \\
        & B1500 & $1500$ & $256^3$ \\
  \tableline
  Set 2 & B936lo & $936$ & $256^3$ \\
        & B936me & $936$ & $512^3$ \\
        & B936hi & $936$ & $1024^3$

  \enddata
\end{deluxetable}

For a successful run of the parallel version of \AHF, five parameters
need to be chosen correctly, each potentially affecting the performance
and reliability of the results.  These are: 
\begin{itemize}
  \item \DomGrid: the size of the domain grid (cf.~\S\ref{subsec:domgrid})
  \item \DomRef: the refinement criterion on the domain grid
(cf.~\S\ref{subsec:domgrid})
  \item \RefRef: the refinement criterion on the refined grid
(cf.~\S\ref{subsec:refcrit})
  \item \LB: the size of boundary zones (cf.~\S\ref{subsec:test_parallel})
  \item \CPU: the number of \MPI-processes used for the analysis
(cf.~\S\ref{subsec:test_parallel})
\end{itemize}
The latter two apply for parallel runs only.  We systematically varied
those parameters and analyzed the three simulations of set~1 described above.  In
table~\ref{tab:testparam}, we summarize the performed analyses.

From each analysis, we will construct the mass-function $N(\Delta M)$
in logarithmic mass bins and the spin parameter distribution $N(\Delta
\lambda)$, where the spin parameter is calculated as \citep{Peebles1969}
\begin{equation}
  \lambda = \frac{J \sqrt{\left| E \right|}}{G M^{5/2}}
\end{equation}
where $J$ is the magnitude of the angular momentum of material within
the virial radius, $M$ is the virial mass and $E$ the total energy of
the system.  The spin parameter $\lambda$ hence combines mass and
velocity information and therefore allows for a good stability test of
the results.  We will further cross-correlate the halo-catalogs to
investigate differences in the deduced individual halo properties.

\begin{deluxetable}{lccccc}
  \tablecaption{Summary of the analyses parameters of B20, B50 and B1500. \label{tab:testparam}}
  \tablehead{
     \colhead{Name\tablenotemark{a}} &
     \colhead{\DomGrid} & 
     \colhead{\CPU\tablenotemark{b}} &
     \colhead{\LB\tablenotemark{c}} &
     \colhead{\DomRef} &
     \colhead{\RefRef}
  }
  \startdata
   064-01-5-5.0-5.0 & $64$ & $1$ & $5$ & $5.0$ & $5.0$ \\
   064-01-5-1.0-5.0 & $64$ & $1$ & $5$ & $1.0$ & $5.0$ \\
   128-01-5-5.0-5.0 & $128$ & $1$ & $5$ & $5.0$ & $5.0$ \\
   128-01-5-1.0-5.0 & $128$ & $1$ & $5$ & $1.0$ & $5.0$ \\
   128-01-5-1.0-4.0 & $128$ & $1$ & $5$ & $1.0$ & $4.0$ \\
   128-01-5-1.0-4.0 & $128$ & $1$ & $5$ & $1.0$ & $3.0$ \\
   128-01-5-1.0-4.0 & $128$ & $1$ & $5$ & $1.0$ & $2.0$ \\
   128-02-4-5.0-5.0 & $128$ & $2$ & $4$ & $5.0$ & $5.0$ \\
   128-02-5-5.0-5.0 & $128$ & $2$ & $5$ & $5.0$ & $5.0$ \\
   128-02-6-5.0-5.0 & $128$ & $2$ & $6$ & $5.0$ & $5.0$ \\
   128-02-7-5.0-5.0 & $128$ & $2$ & $7$ & $5.0$ & $5.0$ \\
   128-04-4-5.0-5.0 & $128$ & $4$ & $4$ & $5.0$ & $5.0$ \\
   128-04-5-5.0-5.0 & $128$ & $4$ & $5$ & $5.0$ & $5.0$ \\
   128-04-6-5.0-5.0 & $128$ & $4$ & $6$ & $5.0$ & $5.0$ \\
   128-04-7-5.0-5.0 & $128$ & $4$ & $7$ & $5.0$ & $5.0$ \\
   128-08-4-5.0-5.0 & $128$ & $8$ & $4$ & $5.0$ & $5.0$ \\
   128-08-5-5.0-5.0 & $128$ & $8$ & $5$ & $5.0$ & $5.0$ \\
   128-08-6-5.0-5.0 & $128$ & $8$ & $6$ & $5.0$ & $5.0$ \\
   128-08-7-5.0-5.0 & $128$ & $8$ & $7$ & $5.0$ & $5.0$ \\
   128-16-4-5.0-5.0 & $128$ & $16$ & $4$ & $5.0$ & $5.0$ \\
   128-16-5-5.0-5.0 & $128$ & $16$ & $5$ & $5.0$ & $5.0$ \\
   128-16-6-5.0-5.0 & $128$ & $16$ & $6$ & $5.0$ & $5.0$ \\
   128-16-7-5.0-5.0 & $128$ & $16$ & $7$ & $5.0$ & $5.0$

  \enddata
  \tablenotetext{a}{The name is constructed from the parameters in the
following way: \textsf{DomGrid-CPU-LB-DomRef-RefRef}.}
  \tablenotetext{b}{Analyses employing one CPU have been performed on an
Opteron running at 1.8GHz, whereas the analyses utilizing more than one
CPU were performed on Xeons running at 3GHz.}
  \tablenotetext{c}{Note that this parameter has no effect for runs with
only one CPU.}
\end{deluxetable}

\subsection{Numerics}
\label{subsec:numerics}

\begin{figure}
  \includegraphics[width=84mm]{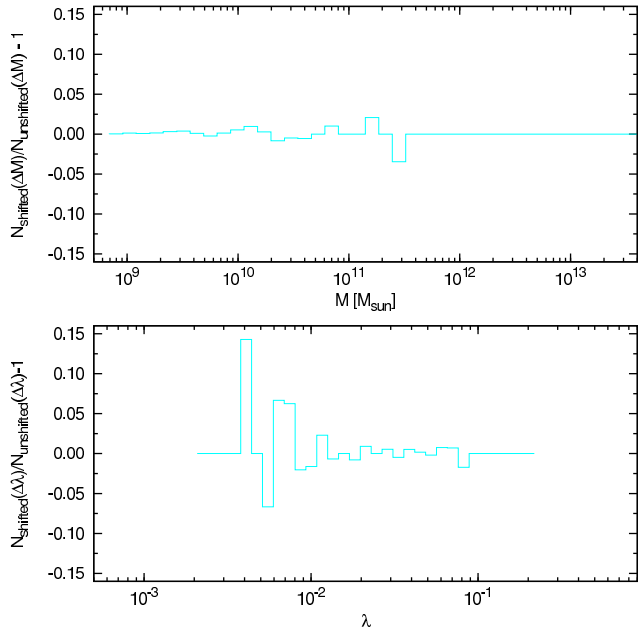}
  \caption{We show a comparison of derived global properties for the
           original particle distribution and a distribution that has
           been shifted by half a domain grid cell ($\DomGrid = 64$) in
           each direction. The ratio of the shifted to the unshifted
           mass function (spin parameter distribution) is shown in the
           upper (lower) panel.}
  \label{fig:shift_global}
\end{figure}

As the grid hierarchy plays the major role in identifying halos, we
test the sensibility of the derived properties on the grid structure by
introducing small numerical artifacts which can lead to slightly
different refinement structures.  To do this, we employ only one
CPU and perform two analyses, one on the original particle distribution
and one on a shifted (by half a domain grid cell in each direction,
taking periodicity into account) particle distribution; we otherwise
keep all other parameters constant at $\DomGrid = 64$, $\DomRef = 5.0$
and $\RefRef = 5.0$.  The ratio of the resulting mass-functions and
spin-parameter distributions are shown in figure~\ref{fig:shift_global}.

  Small deviations can be seen in the mass function, which is however of
the order of $1-3$ per cent in a few bins and then mostly due to one or
two haloes changing a bin.  This translate into some scatter in the spin
parameter distribution, however the seemingly large deviation on the
low- and high-spin end are artificially enhanced due to the low number
counts in the bins, we also excluded halos with less than 100 particles
(cf.\ \S\ref{subsec:refcrit}) from this plot.  The scatter here is also
of the order of $1-3$ per cent.  It is important to keep in mind that
the derived properties can vary to this extent simply due to numerical
effects.

However, the main point of this comparison is to verify that, up to
numerical effects caused by the perturbed density field, we do not
introduce any systematic deviation in the statistical properties.

\subsection{Domain Grid}
\label{subsec:domgrid}

\begin{figure*}
  \includegraphics[width=169mm]{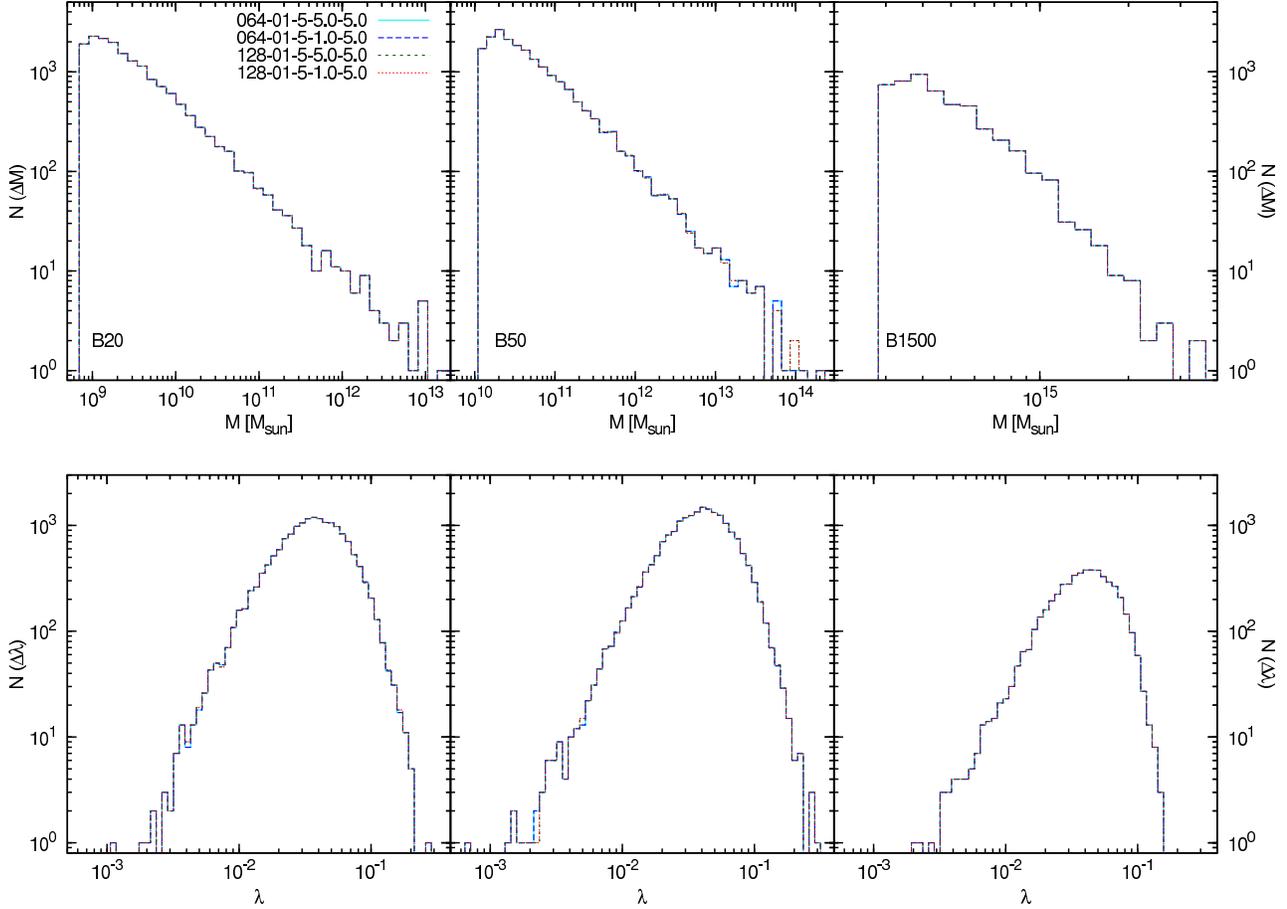}
  \caption{Here we show the mass-function (upper row) and the spin
parameter distribution (lower row) for the three simulations.  The left
(middle, right) column corresponds to the B20 (B50, B1500),
respectively.  We employed one CPU and varied the domain grid between
$64$ and $128$ and for each choice also varied the refinement criterion
on the domain grid between $5.0$ and $1.0$, keeping the refinement
criterion on the refined grids fixed at $5.0$.}
  \label{fig:serial_dom}
\end{figure*}

We now focus on the choice of \DomGrid, i.e. the size of the domain
grid.  To this extent, we employ one process and vary the domain grid
and the refinement criterion \DomRef\ on the domain grid.  In
figure~\ref{fig:serial_dom} we show the deduced mass-function and spin
parameter distribution for the four cases.  As can be seen, the impact
of the domain grid choice is negligible;  small deviations can be seen
at the high mass end of the mass function in B50, however these
are caused by (of order) one halo changing bin across runs with varied
parameters.  We will discuss the drop of $N(\Delta M)$ at the low
$M$ end of the mass function below.

In view of the parallel strategy the insensitivity of the results to the
choice of the domain grid is reassuring, as we can use a rather coarse domain
grid and start the refinement hierarchy from there; note that the domain
grid will be allocated completely on each CPU and hence choosing a fine
grid would lead to a large number of empty cells in the parallel
version.

It can also be seen that the choice of the refinement criterion on the
domain grid has no impact.  This is because the domain grid is coarse
enough as to justify refinement everywhere anyhow.  In fact, only very
pronounced underdense regions would not trigger refinements for the
choices of parameters and particle resolution.

\subsection{Refinement Criterion}
\label{subsec:refcrit}

Now we investigate the effect of choosing a different refinement
criterion (\RefRef) for the refined grids.  We limit ourselves to a
domain grid of $\DomGrid = 128$ cells per dimension and use a refinement
criterion on the domain grid of $\DomRef = 1.0$ with one process.  We
then vary the refinement criterion on the refined grids from $\RefRef =
5.0$ (this corresponds to analysis already used above when investigating
the impact of the domain grid and its refinement criterion) to $\RefRef
= 2.0$ in steps of $1.0$.

\begin{figure*}
  \includegraphics[width=169mm]{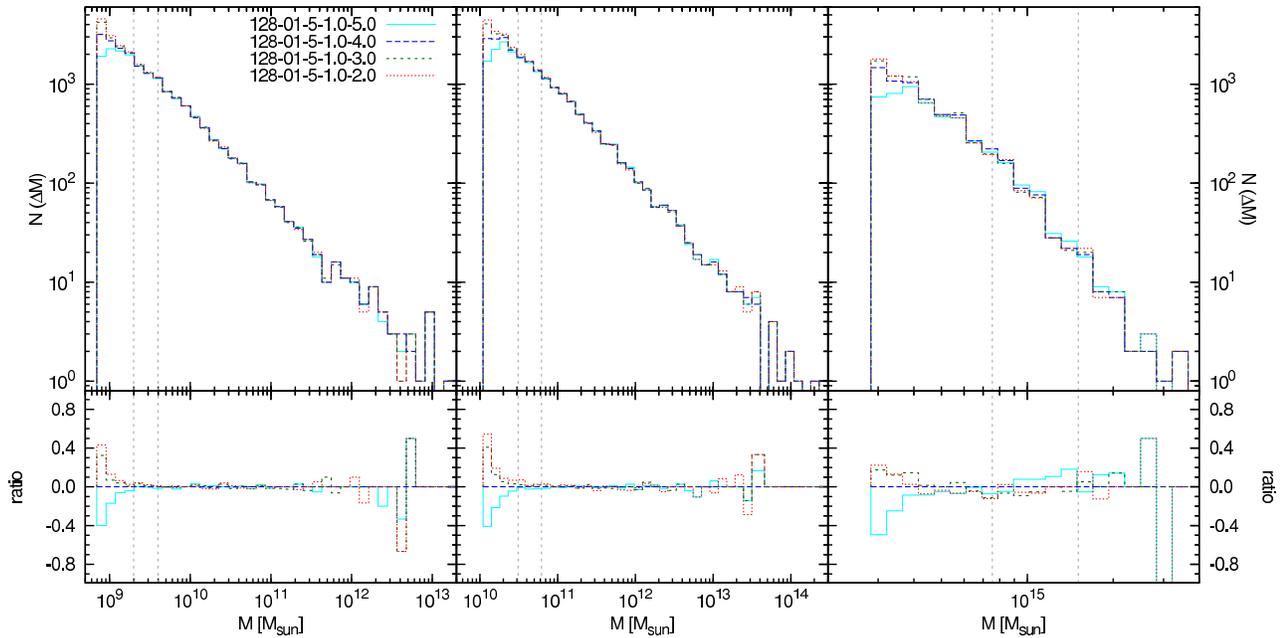}
  \caption{The effect of the refinement criterion on the refined grids
on the mass function.  We show the mass functions derived for the
given analyses in the top and also deviations arbitrarily normalized to the
128-01-1.0-4.0 in the smaller bottom panels.  For guidance, the vertical
lines indicate the mass corresponding to halos with 50 and 100
particles, respectively.  The columns are organized as in
figure~\ref{fig:serial_dom}.}
  \label{fig:serial_mass_ref}
\end{figure*}

\begin{figure*}
  \includegraphics[width=169mm]{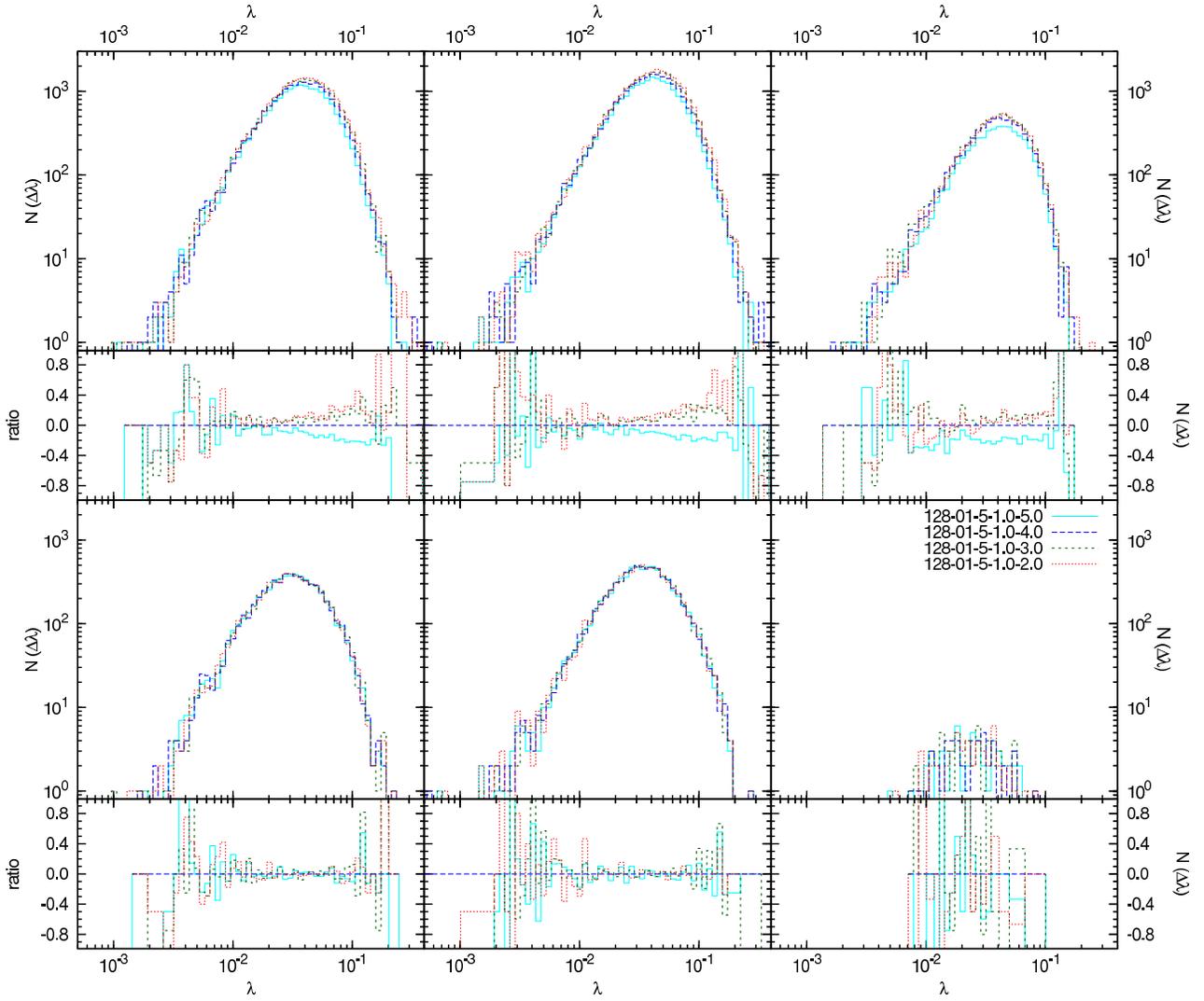}
  \caption{As figure~\ref{fig:serial_dom} but for the spin parameter
distribution of the whole sample of halos (top) and restricted to
halos resolved with more than 100 particles (bottom).}
  \label{fig:serial_lambda_ref}
\end{figure*}

In figures~\ref{fig:serial_mass_ref} and
\ref{fig:serial_lambda_ref} we show the effect of varying the
refinement criterion on the refined grids on the mass-function and the
spin-parameter distribution.  It can be seen that the mass-function 
changes mainly at the low mass end.  The changes are  related to the
mass discreteness: The more particles a halo is composed of, the easier
it is to pick it up with our refinement criterion based on the number of
particles within a cell.  Hence it can be readily understood that by
forcing a smaller criterion, more halos with a small amount of
particles will be found.

This is important for the completeness of the deduced halo
catalogs: only when choosing a very small refinement criterion
($\lesssim 3.0$), we are complete at the
low-particle-count\footnote{This translates in general into a low mass
halo, however, the important quantity is really the number of particles,
not their mass.  In zoomed simulations, low-mass halos in the zoomed
region might be found completely whereas halos of the same mass in the
low resolution regions are not picked up.} end.

The spin parameter distribution is also affected.  We can see in the
top row of figure~\ref{fig:serial_lambda_ref} that the peak of the
distribution shifts slightly and is reduced in height\footnote{The
latter is due to the fact that the distribution is not normalized.}.
However, when only including halos with more than $100$ particles
(shown in the bottom panels of figure~\ref{fig:serial_lambda_ref}) in
the calculation of the distribution, the shift is reduced and the
distributions coincide; note that the most massive halo in the analyses
of B1500 is resolved with only 237 particles.

As we will discuss later, the choice of the refinement criterion
severely impacts on the runtime and the memory requirements (see
figure~\ref{fig:scaling_refref}) and hence should be lowered only with care.

\subsection{The boundary zone and the number of processes}
\label{subsec:test_parallel}

We will now investigate the effect of the size of the boundary zone and
the number of processes.  As the reference model we use a serial run
with a domain grid of $\DomGrid = 128$ cells per dimension, a refinement criterion
of $\DomRef = 1.0$ on the domain grid and a refinement criterion of
$\RefRef = 5.0$ on the
refinements; note that the \LB\ parameter has no effect in the case of
a serial run.  We also change the number of processes involved in the
analysis from $1$ (the reference analysis for each box) in factors of
$2$ to $16$, in which case the \LB\ parameter has a very significant
meaning:  Recall that the volume decomposition scheme not only assigns
to each CPU the associated unique volume, but also a copy of the
boundary layer with a thickness of $1$ cell.  We increase the size of
the decomposition grid from a $2^4=16$ ($\LB = 4$) cells per dimension
grid by factors of two up to a $2^7=128$ ($\LB = 7$) cells per dimension
grid.  To correctly identify a halo located very close to a boundary,
the buffer volume must be large enough to encompass all its particles,
namely the thickness of the boundary should be
$R_\mathrm{vir}^\mathrm{max}$ (cf. equation \ref{eq:boundary}). This
condition is violated for B20 from $\LB > 4$ on and for B50 for $\LB >
5$, whereas in B1500 it would only be violated for $\LB > 8$.  Hence we
expect to see differences in the derived properties due to the volume
splitting for all analyses violating this condition.

\subsubsection{Statistical comparison}
\label{subsubsec:statistical}

\begin{figure*}
  \includegraphics[width=169mm]{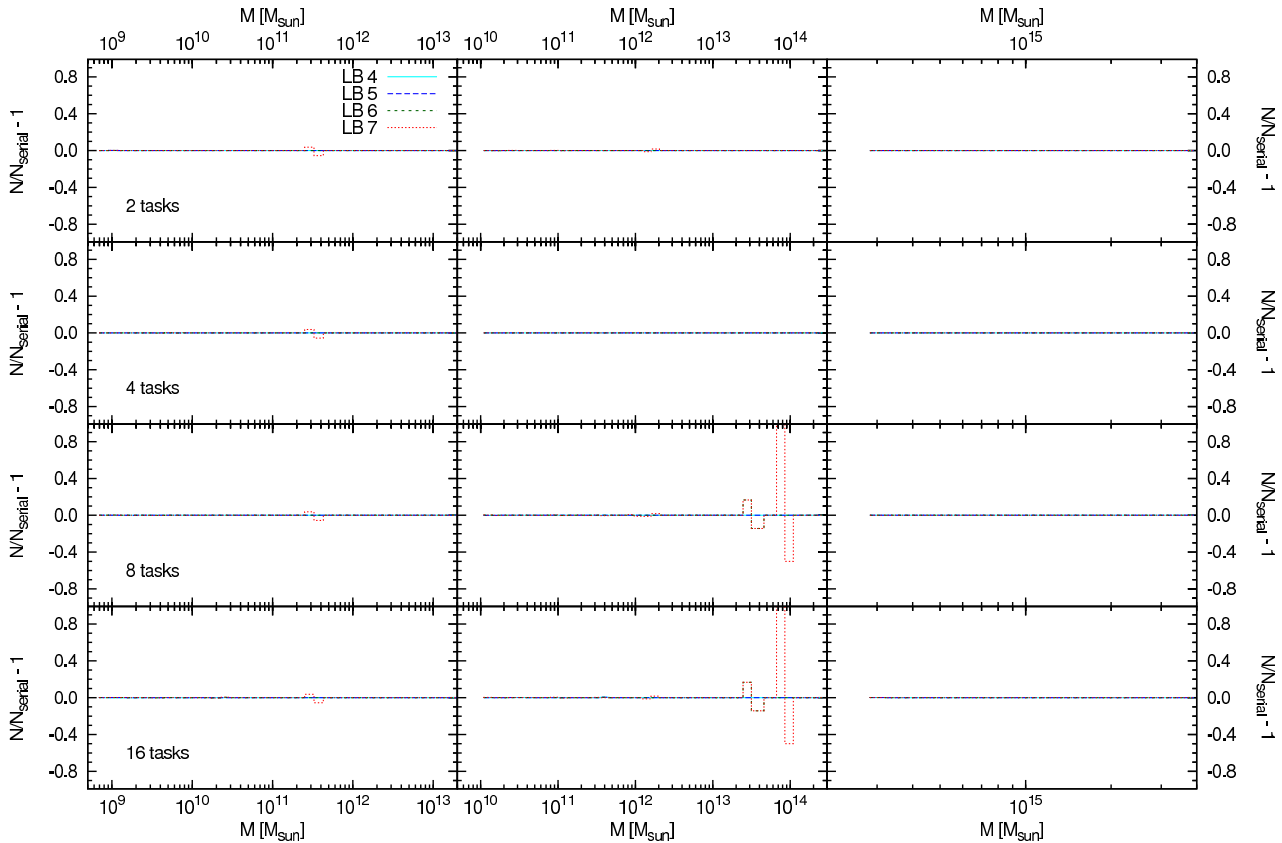}
  \caption{We show the massfunctions derived for our three simulations
           varying the number of processes and size of the domain
           decomposition grid.}
  \label{fig:parallel_massfunc}
\end{figure*}

\begin{figure*}
  \includegraphics[width=169mm]{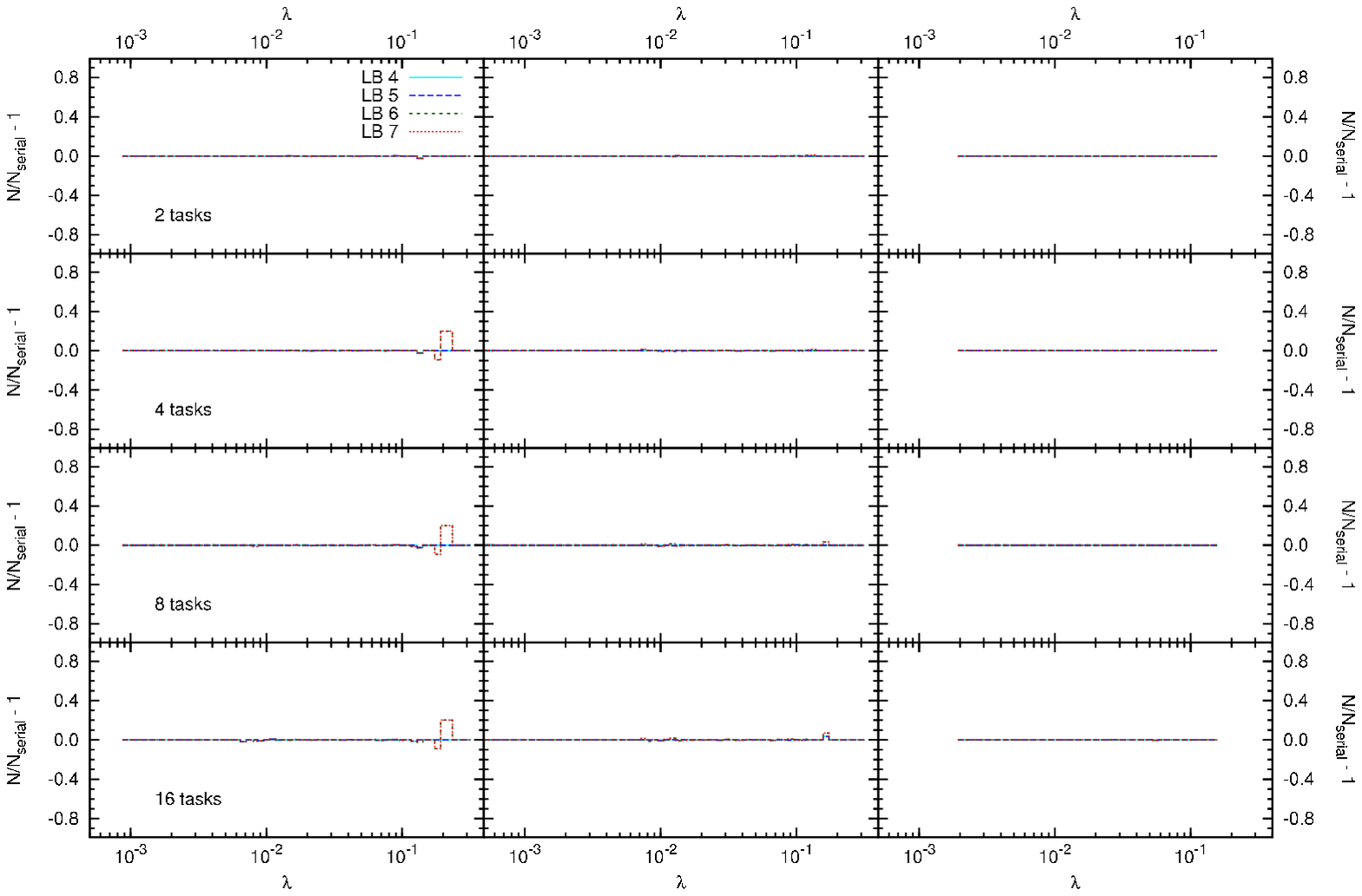}
  \caption{As figure~\ref{fig:parallel_massfunc} but for the spin
           parameter distribution.}
  \label{fig:parallel_lambda}
\end{figure*}

We first concentrate on integrated properties, namely the mass functions
and the spin parameter distributions again.  In
figure~\ref{fig:parallel_massfunc} we show the ratio of the mass
functions, whereas in figure~\ref{fig:parallel_lambda} the ratio of the
spin parameter distribution is depicted.  In all cases we have taken the
serial run 128-01-5-5.0-5.0 as the reference and show relative
deviations from it.  In each figure the number of employed processes
increases from top to bottom from $2$ to $16$, whereas each single plot
shows the derived mass function (spin parameter distribution) for the
four choices of the boundary size.

It can be seen, that the integrated properties appear to be in general rather
unaffected by the choice of the number of CPUs or the \LB\ level.  For
B20, the mass function only shows a difference for the \LB 7 analysis,
the B50 shows differences at the high mass end for \LB 6 and \LB 7 and more
than $4$ CPUs. Looking at the spin parameter distribution, only B20
shows a difference at the large $\lambda$ end for \LB 6 and \LB 7.
B1500 is completely unaffected by any choice of tested parameters.  As
we have alluded to above, this is expected and we can clearly see two
effects coming into play:  First, the smaller the boundary zone is, the
higher the probability that a halo might not be available entirely to the
analysing task.  Second, increasing the number of CPUs simultaneously
increases the amount of boundary cells and hence the chance to provoke
the aforementioned effect.

However, it is interesting to note that in a statistical view we really
have to go to the worst case choices (large \LB, many CPUs), to get a
significant effect.  Also we should note that the shown deviations are
relative and are hence more pronounced in the bins with low counts,
namely the outskirts of the spin parameter distribution and the high
mass end of the mass function.  Choosing the correct \LB\ value, we can
see that the integrated properties to not depend on the number of CPUs
involved.

\subsubsection{Halo-halo cross comparison}
\label{subsubsec:halohalocrosscomp}

We will now investigate the dependence of the individual halo properties
on the parallelizing parameters.  To this extent we perform a
cross-correlation of particles in identified objects between the
different parallel runs and -- taken as the reference -- a serial run.

We employ a tool included in the \AHF-distribution which establishes for
two given halo catalogs --- for this purpose consisting of the particle
IDs --- for each halo from the first catalog the best matching halo from
the second catalog.  This is done by maximizing the quantity
\begin{equation}
  \xi = \frac{N^2_{\mathrm{shared}}}{N_1 N_2}
\end{equation}
where $N_{\mathrm{shared}}$ is number of shared particles between the two
halos and $N_1$ and $N_2$ are the total number of particles of the halo
from the first and the second catalog, respectively.

In the ideal case, we expect every particle found in a halo in the
reference analysis to also appear in the corresponding halo in the
parallel version.  Due to the choice of the boundary size, this might
not be the case, as for halos located close to a boundary the required
particles might not be stored in the boundary.  Also numerical effects
from a different grid structure in the boundary can lead to slight
variations in the identified halos, as we have demonstrated already in
figure~\ref{fig:shift_global}.  Additionally we should note that the
catalogs produced by two runs with different number of CPUs cannot be
compared line by line, as, even though the ordering is by number of
particles, halos with the same number of particles are not guaranteed to
appear in the same order.

\begin{figure*}
  \includegraphics[width=169mm]{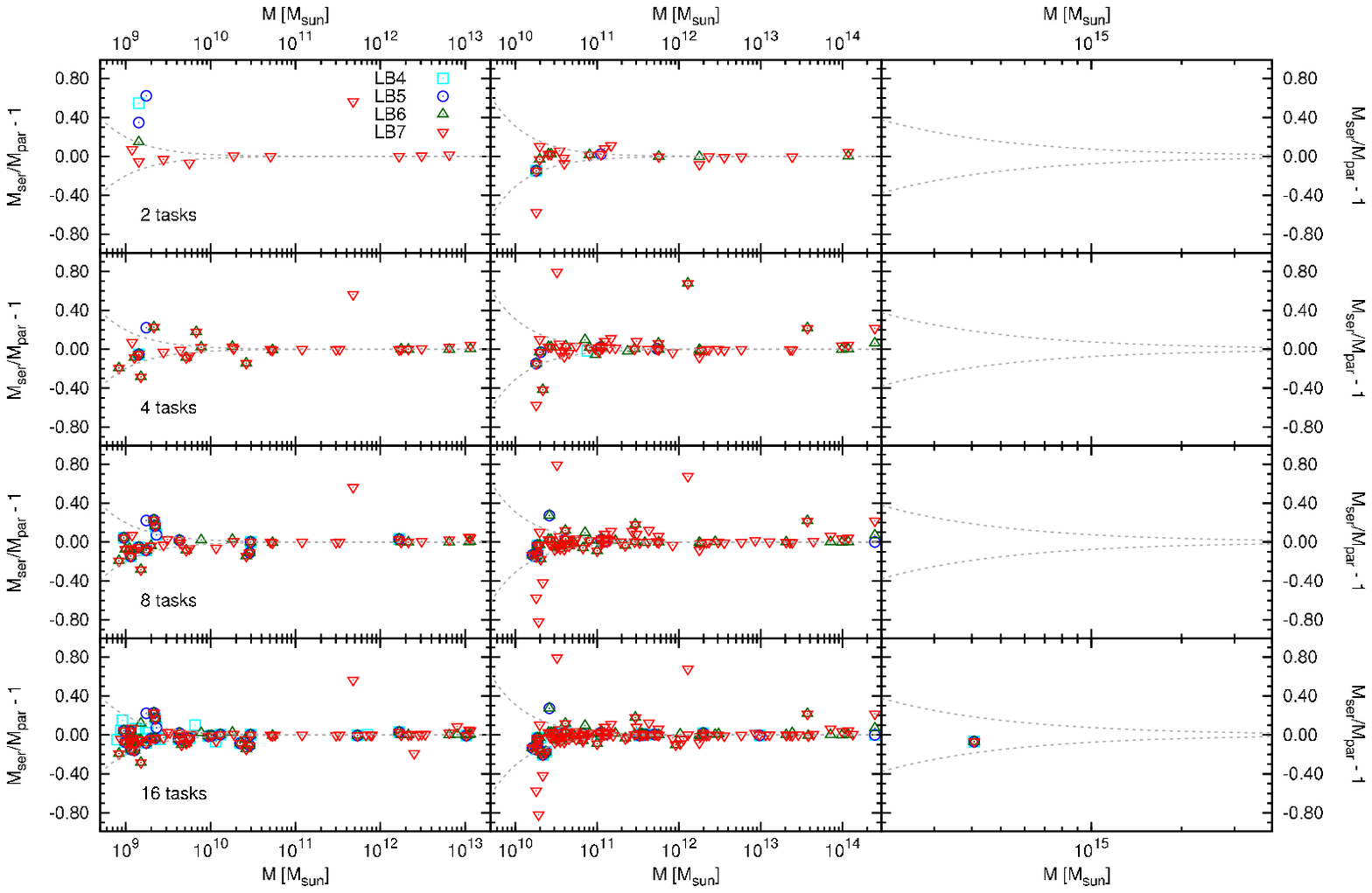}
  \caption{Mass ratio of halos in the parallel version using 2 (4, 8,
16) processes from top to bottom in the three boxes. Open boxes
(circles, upright triangles, downright triangles) are used to
indicate \LB 4 (5, 6, 7).  The additional curved lines correspond
to a difference in halo mass of $\pm5$ particles. Note that we only show
those halos that have a ratio not equal to unity and we see the expected
behaviour that the mismatch becomes more promiment in the case of a
large \LB\ (cf.\ discussion in \S\ref{subsubsec:halohalocrosscomp}.}
  \label{fig:parallel_ccor_mass}
\end{figure*}

\begin{figure*}
  \includegraphics[width=169mm]{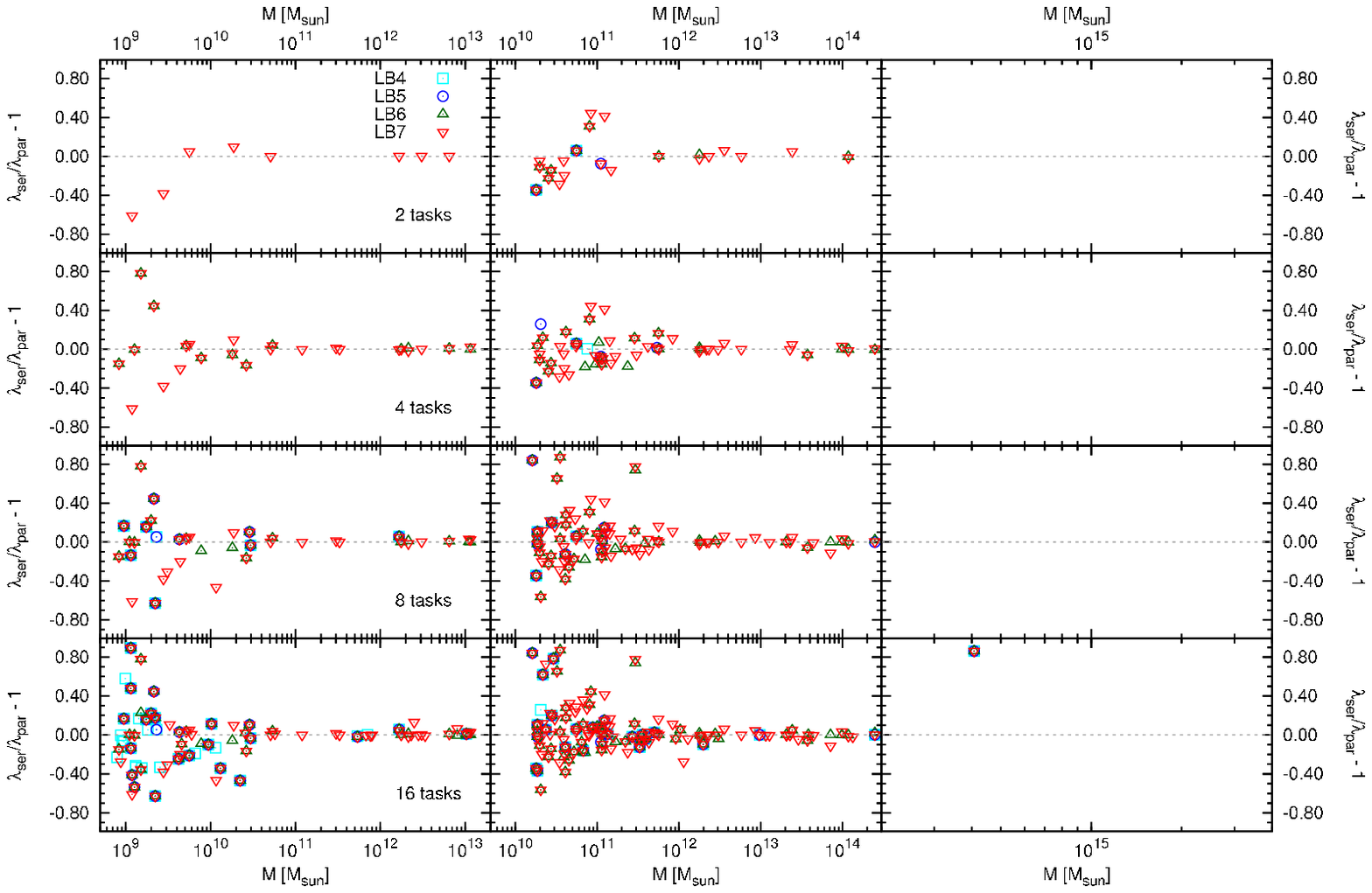}
  \caption{As figure~\ref{fig:parallel_ccor_mass} but for the spin
parameter as a function of halo mass.}
  \label{fig:parallel_ccor_lambda}
\end{figure*}

We show this impact for the mass of the halos in
figure~\ref{fig:parallel_ccor_mass} and for the spin parameter in
figure~\ref{fig:parallel_ccor_lambda}.  Note that we only plot data
points for halos that have different properties.  In total we have
$16631$ ($19621$, $4971$) objects in B20 (B50, B1500).

It is interesting to note that even though the distribution of
integrated properties (cf. \S\ref{subsubsec:statistical})
show no significant effect, we do find slightly different results for
individual halos.  However, the observed differences in B20 and B50
for \LB 6 and \LB 7 are to be expected, as in these cases the
boundary zones are to thin to cope with the extent of the halos, as we
have alluded to above.  For B1500 we would need to increase the \LB\
level to a significantly larger number to provoke the same effects.
Additionally the differences found are generally of the order of a few
particles as indicated by the dashed lines in
figure~\ref{fig:parallel_ccor_mass} which correspond to a difference of
$5$ particles.  We have observed in \S\ref{subsec:numerics} that
numerical noise --- as introduced by shifting the particle positions ---
already can induce fluctuations of the order of a few particles (cf.
figure~\ref{fig:shift_global}).

Generally, however, we can say that, choosing a sane \LB\ level
complying with equation~\ref{eq:boundary}, the
parallel version gives the same result as the serial version.

\section{Comparison to other halo finders and theortical predictions}
\label{sec:comparison}

In this section we compare the results of \AHF\ to three other halo
finding mechanisms.  We restrict ourselves to the analysis of B20 and
use a \FOF\ halo finder \citep{Davis1985}, \SKID\
\citep{PhD_Stadel2001}, and an implementation of the \BDM\
\citep{Klypin1997} method.

For the \FOF\ run, we use a linking length of $0.17$, which yields an
overdensity at the outer radius comparable to the virial overdensity
used in \AHF.  In the \BDM\ analysis, in order to find (potential) halo
centers we used a density field smoothed on a scale of $10 h^{-1}
\mathrm{kpc}$, approximately corresponding to a mass scale of $10^9
h^{-1} \mathrm{Mpc}$, and we therefore expect the mass function to not
be complete on this scale.  Otherwise we use a variant of the original
\BDM\ scheme to identify the final halos and their properties
\citep[cf.\ Appendix~B in][]{Bullock2001}.  For the \SKID\ run, we used
a linking length of $3\epsilon$, where $\epsilon$ is the force
resolution of the simulation ($\epsilon = 1.5 h^{-1} \mathrm{kpc}$).

Besides a direct comparison between these halo finders in
\S\ref{subsec:hfcomp}, we will further investigate in
subsection~\ref{subsec:theomassfunc} how well theoretical descriptions
presented in the literature describe the numerically obtained mass
functions.  To this extent, we use the highest resolved simulation of
set 2, B963hi.
 
As one of our claims is that \AHF\ is capable of identifying
field halos as well as subhalos simultaneously it also appears
mandatory to compare the derived subhalo mass function against the
findings of other groups. These results will be presented in
subsection~\ref{subsec:submassfunc}.

\subsection{Comparison to other halo finders}
\label{subsec:hfcomp}

\begin{figure}
  \includegraphics[width=84mm]{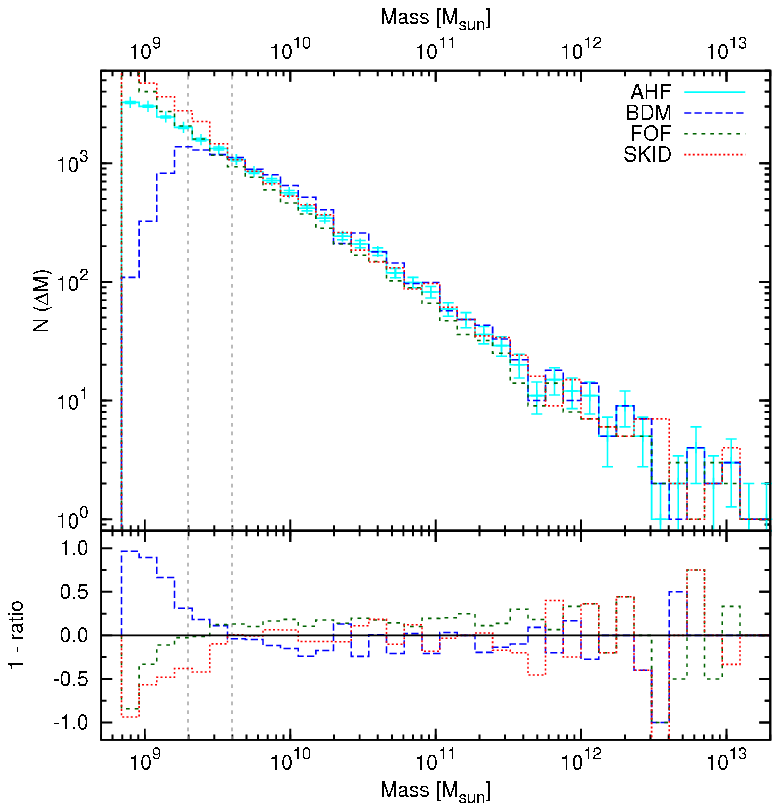}
  \caption{The mass function derived with different halo finders for
B20.  The top graph depicts the actual mass function and we only show
the Poisson errors for the \AHF\ data for clarity. In the lower frame
the ratio of the mass functions to the \AHF\ mass function is shown.
Additionally, the dashed vertical lines indicate the halo masses that
correspond to 50 and 100 particles, respectively.}
  \label{fig:hfcomp_mass}
\end{figure}

In figure~\ref{fig:hfcomp_mass} we show the derived mass functions for
the four differnt codes in the top panel alongside Poissonian error
bars for each bin.  The lower panel investigates the relative
deviation of the three additional halo finders tested in this section
to the results derived from our \AHF\ run.  To this end, we calculate
\begin{equation}
 \frac{N_{\AHF}(\Delta M) - N_X(\Delta M)}{N_{\AHF}(\Delta M)}
\end{equation}
where $X$ is either \BDM, \FOF\ or \SKID.  Hence a positive deviation
means that the \AHF\ run found more halos in the given bin than the halo
finder compared to.

  Please note that with the settings used, we do not consider the mass
function below masses corresponding to 100 particles to be complete in
the \AHF\ analysis.  Also the sharp decline of the \BDM\ function is not
due to the method but rather to the smoothed density field alluded to
above.

  Generally, we find the mass functions derived with \BDM\ and \SKID\ to
match the results from the \AHF\ run within the error bars quite well,
whereas the \FOF\ analysis shows a systematic offset.  This mismatch
between mass functions deduced with the \FOF\ method and \SO\ based
methods is known and expected \citep[see, e.g., the discussion
in][concerning the relation of \FOF\ masses to \SO\
masses]{Tinker2008, Lukic2009}.

\subsection{Comparison to theoretical mass functions}
\label{subsec:theomassfunc}
\begin{figure}
  \includegraphics[width=84mm]{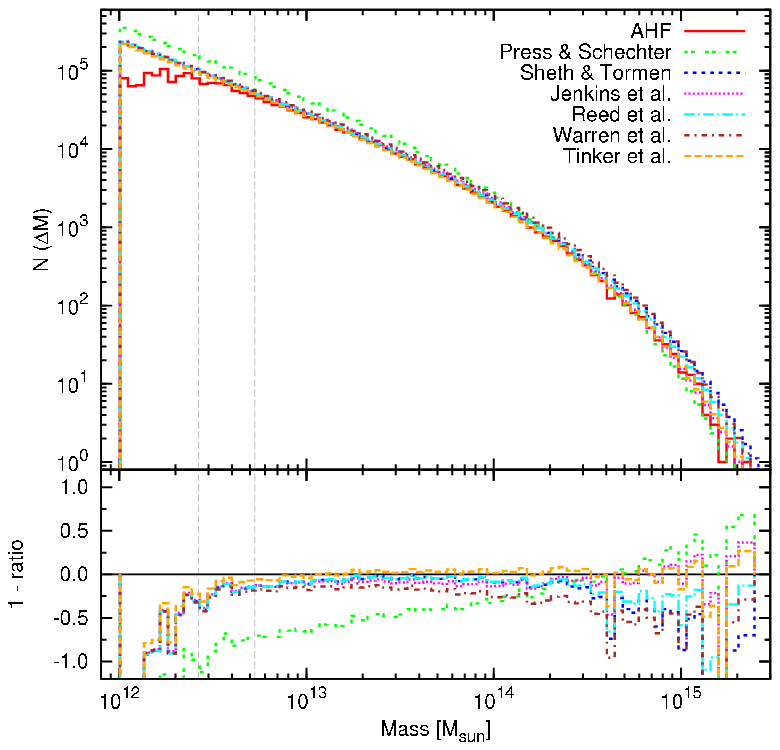}
  \caption{In this figure, the mass function for B963hi derived with
\AHF\ is compared to a variety of theoretical forms. Again, the vertical
dashed lines correspond to the halo masses of 50 and 100 particles,
respectively.}
  \label{fig:hfcomp_mass_high}
\end{figure}

The entering of the era of precision cosmology in the last couple of
years made it possible to derive the mass function of gravitationally
bound objects in cosmological simulations with unprecedented
accuracy. This led to an emergence of refined analytical formulae
\citep[cf. ][]{Sheth1999, Jenkins2001, Reed2003, Warren2006,
Tinker2008} and it only appears natural to test whether or not our
halo finder \AHF\ complies with these prescriptions. To this
extent we are utilizing our best resolved simulation (i.e.\ B963hi of
Set 2, cf. Table~\ref{tab:simulations}) and present in
figure~\ref{fig:hfcomp_mass_high} a comparison against the mass
functions proposed by \citet{Press1974, Sheth1999, Jenkins2001,
Reed2003, Warren2006} and \citet{Tinker2008}.

  In this respect it must be noted that the proposed theoretical mass
functions are calibrated to mass functions utilizing different halo
overdensities.  The mass function of \citet{Sheth1999} is based on halo
catalogs applying an overdensity of $\Delta = 178$ \citep{Tormen1998a},
and the \citet{Reed2003} and \citet{Warren2006} functions are formulated
for overdensities of $\Delta = 200$. \citet{Jenkins2001} provide various
calibration and we use their equation~B4 which is calibrated to a
$\Lambda$CDM simulation and an \SO-halo finder with an overdensity of
$\Delta = 324$, quite close to our value of $\Delta = 340$.  The
\citet{Tinker2008} mass function explicitly includes the overdensity as
a free parameter and as such we use the value of our catalog to produce
the mass function.

  We find the \citet{Tinker2008} formula to give an excellent
description of the observed mass function and also the
\citet{Jenkins2001} formula provides a good description, albeit
overestimating the number of halos at intermediate masses, which might
be due to the slightly wrong overdensity.  While the other formulas also
provide an adequate description for the mass function at intermediate
masses, they overestimate the number of halos at the high mass end.  As
has been eluded to above this is to be expected and can be understood by
the different halo overdensities they have been calibrated to.  However,
clearly the \citet{Press1974} prescription does not provide a good
description of the halo mass function in this case.  As the
\citet{Tinker2008} formula provides an excellent fit to our data, we
refer the reader to their work for further discussions concerning the
impact of the different overdensities used to define a halo.

\subsection{The subhalo mass function}
\label{subsec:submassfunc}
As our halo finder simultaneously finds field and sub-halos without
the need to refine the parameters for the latter it appears mandatory
to compare the derived subhalo mass function against results in the
literature, too. To this extent, we identify the substructure of the
most massive halo in B20 with our cross matching tool (cf.\
\S\ref{subsubsec:halohalocrosscomp}).  We restrict ourselves here to
only investigating this particular halo, as the particle resolution
for the other available simulations is not sufficient to resolve the
subhalo population adequately.  Also, we did use the 128-01-5-1.0-2.0
analysis to have a high sensitivity to very small halos
(cf. discussion of \AHF's parameters in section~\ref{sec:testing}).

\begin{figure}
  \includegraphics[width=84mm]{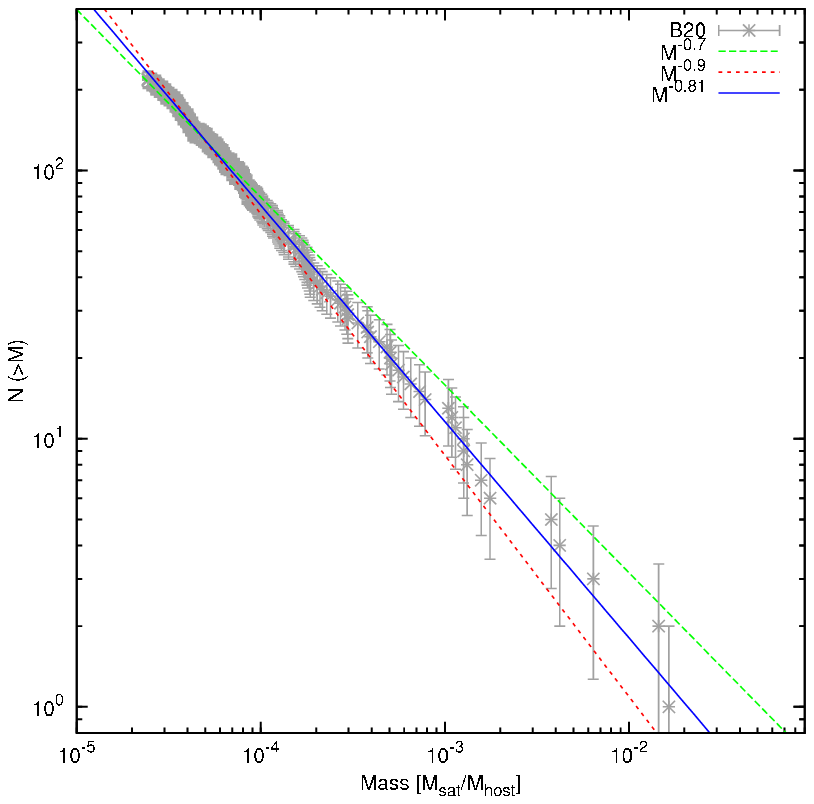}
  \caption{The {\em cumulative} subhalo mass function for the most
massive halo in B20.  This halo is resolved with $\approx 8\times10^5$
particles and contains $221$ resolved subhalos.  Also we show three
power law fits, two with a fixed slope of $-0.7$ and $-0.9$,
respectively, and a third with the slope as a free parameter, yielding
$-0.81$. Note that this plot is based on the 128-01-5-1.0-2.0
analysis.}
  \label{fig:hfcomp_subhalo_theory}
\end{figure}

It has been found before \citep[e.g.][]{Ghigna2000, Helmi2002,
  DeLucia2004, Gao2004, vandenBosch2005, Diemand2007, Giocoli2008,
  Springel2008} that the subhalo mass function can be described with a
functional form $N_{\rm sub}(>M) \propto M^{-\alpha}$, with $\alpha =
0.7 \ldots 0.9$.  In figure~\ref{fig:hfcomp_subhalo_theory} we show
the cumulative mass function of the most massive halo in B20 and
provide -- as guide for the eye -- two power laws with those limiting
slopes.  Additionally, we fitted a power law to the actual $N_{\rm
  sub}^{\rm AHF}(>M)$, yielding a slope of $\alpha = 0.81$.  This test
confirms the ability of \AHF\ to reproduce previous findings for the
subhalo mass function and hence underlining its ability to function as
a universal halo finder.

\section{Results}
\label{sec:results}

In this section we summarize the results obtained from the parameter
study (\S\ref{subsec:res_pdep}) and present the requirements of memory
and time.  We believe this to be important information for understanding
how to maximize the gain from using \AHF.  In this respect, we
specifically investigate the achieved load-balancing of the parallel
version (\S\ref{subsec:res_lb}).  We further present the scaling of
\AHF\ with increasing problem size in \S\ref{subsec:scalability}.

\subsection{Parameter dependence}
\label{subsec:res_pdep}

\begin{figure}
  \includegraphics[width=84mm]{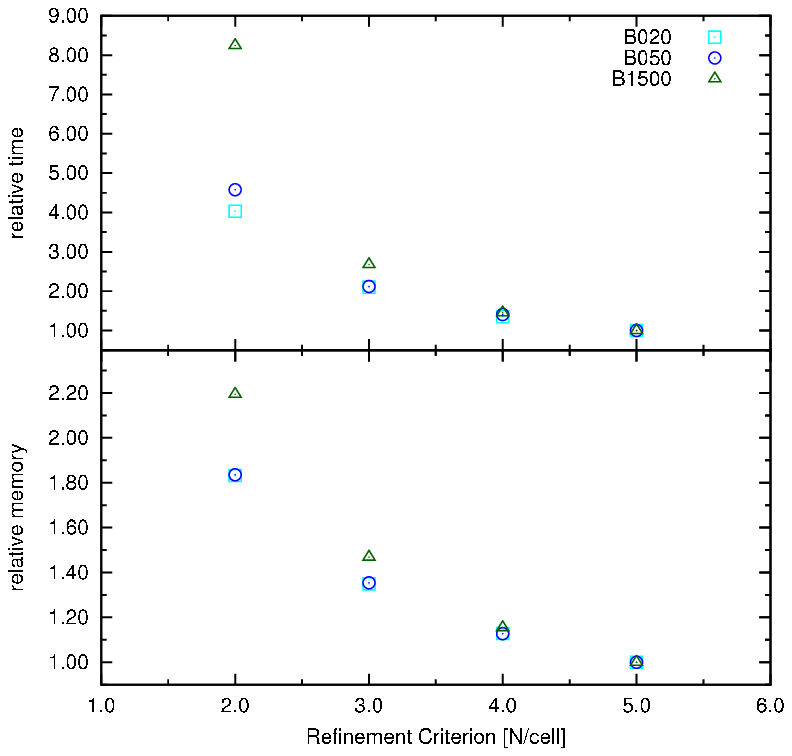}
  \caption{The impact of the refinement criterion on the refined grids
on the memory and time consumption of \AHF.  We normalize the
128-01-5-1.0-X.0 runs to the 128-01-5-1.0-5.0 run and show the relative
time in the upper and the relative memory consumption in the lower
panel.}
  \label{fig:scaling_refref}
\end{figure}

As we have shown, the \DomGrid\ and \DomRef\ parameters have a negligible
impact on the derived properties  and given an adequate choice for the LB
level, also the number of employed CPUs has no impact on the derived
parameters.  However the \RefRef\ parameter can be used to force
completeness of the mass function down to small particle counts.  This
comes with a price as we show in figure~\ref{fig:scaling_refref}; this
figure depicts the increase in run time and memory requirement relative
to a choice of $\RefRef = 5.0$ when changing the \RefRef\ parameter.  As we have
shown in figure~\ref{fig:serial_mass_ref} a \RefRef\ parameter of $3.0$
would be required to achieve completeness for halos with less than $50$
particles.  However this is a factor of $2-3$ in runtime and an increase
in memory requirement by $\sim40$ per cent.  These numbers are derived
from a serial run.

\subsection{Loadbalancing}
\label{subsec:res_lb}

We will now investigate the quality of the load-balancing of \AHF.
All parallel analyses of set~1 (cf.~table~\ref{tab:simulations}) have
been performed on the \textsc{damiana}\footnote{For more details on the
topology of the cluster, see here:
\url{http://supercomputers.aei.mpg.de/damiana/}} cluster, where each
node is equipped with two Dual Core Xeon processors running at $3$GHz.
Due to memory constraint --- especially for the variation of the
\RefRef\ parameter --- the serial analyses have been performed on one of
our local machines (Opteron running at $1.8$GHz) where we were certain
to not be bound by memory.  For this reason, the times between the
serial and parallel runs are not directly comparable.

\begin{figure*}
  \includegraphics[width=169mm]{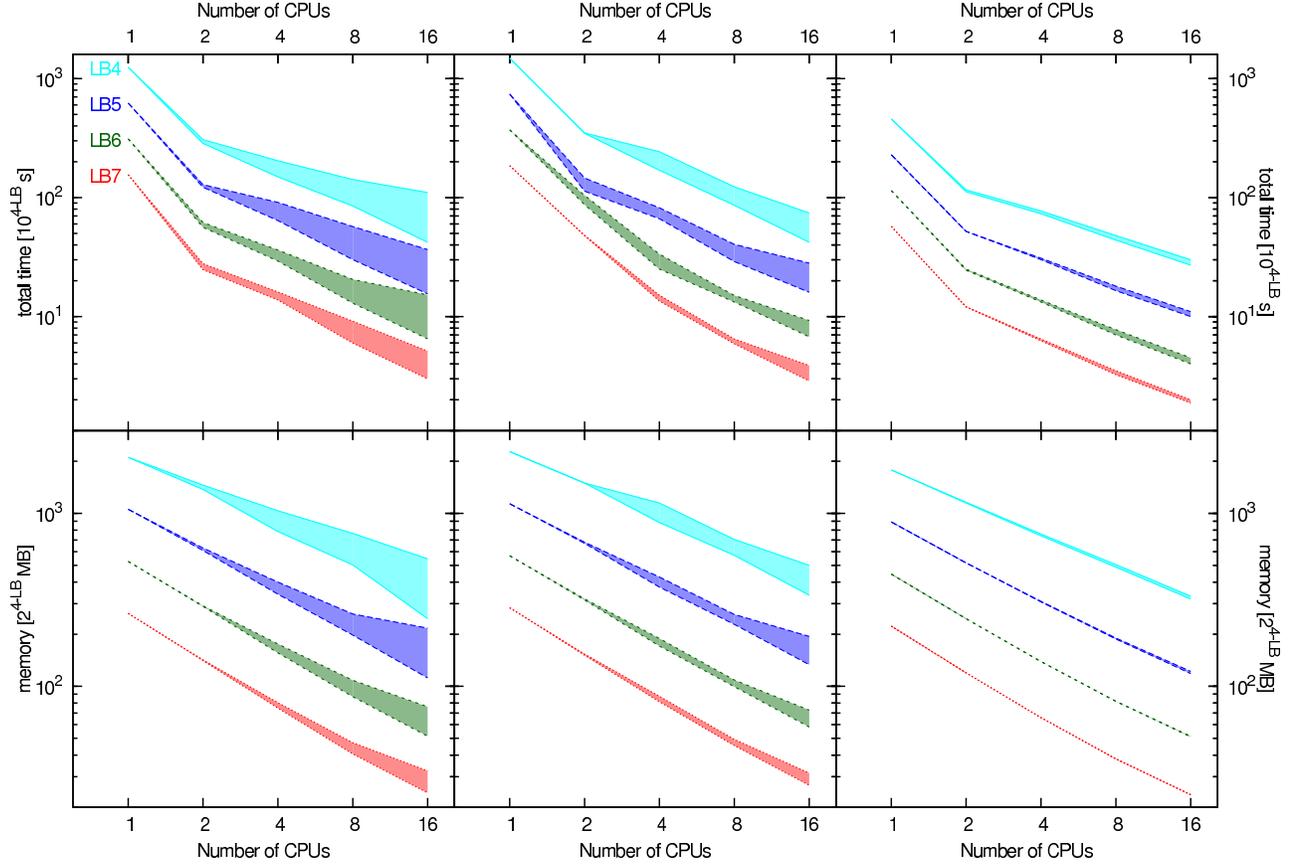}
  \caption{We show the absolute times required for the different
analyses in the top row and the total memory requirement for the grid
hierarchy and the particles in the lower row.  The columns are ordered
as described in figure~\ref{fig:serial_dom}.  The top row gives the
duration for the whole run, whereas the two middle rows give the times
needed for the grid generation and the actual halo analysis.  The last
row shows the amount of memory required to store the particle
information and the grid hierarchy.  We use the 128-01-5-5.0-5.0 run for
producing the data point for 1 CPU and plot for the parallel runs the
maximal and minimal values.  Each panel shows the results for different
choices of LB and for clarity, the curves are offset to one another by
factors two.  Note that runs with one CPU were, due to memory
restrictions, performed on a different, slower, machine than the
parallel runs, hence the kink in the timing curves at 2 CPUs is not
real.  However the memory curves are not affected by this.}
  \label{fig:scalability_abs}
\end{figure*}

In figure~\ref{fig:scalability_abs} we show the absolute times of the runs
and we give the span of run times of the separate processes for the
parallel analyses.
It can be seen that increasing
the number of CPUs involved in the analysis indeed reduces the runtime.
However, when
increasing the number of CPUs a widening of the spread between fastest
and the slowest task can be observed.  This is most pronounced in B20
and in particular for the time required to do the halo analysis, whereas
the time required to generate the grid hierarchy exhibits a smaller spread.
The large spread can also be seen in the memory requirement for the
grids, shown in the lower row of figure~\ref{fig:scalability_abs}.  This
is especially pronounced for small \LB\ levels.  When going to larger
boxes the spread becomes smaller. 

The behaviour of increasing spread with decreasing boxsize can be
readily understood as the effect of the non-homogeneity of the particle
distribution manifesting itself.  A small box will typically be
dominated by one big halo, whereas in large boxes there tend to be
more equally sized objects.  As we do not yet allow one single halo to
be jointly analysed by more than one CPU, the analysis of the most
massive halo will basically dictate the achievable speed-up.  Of course,
when decreasing the boundary size (increasing the \LB\ level), the spread
can be reduced, however, the results will be tainted as has been shown
in figures~\ref{fig:parallel_ccor_mass} and
\ref{fig:parallel_ccor_lambda}.

\begin{figure*}
  \includegraphics[width=169mm]{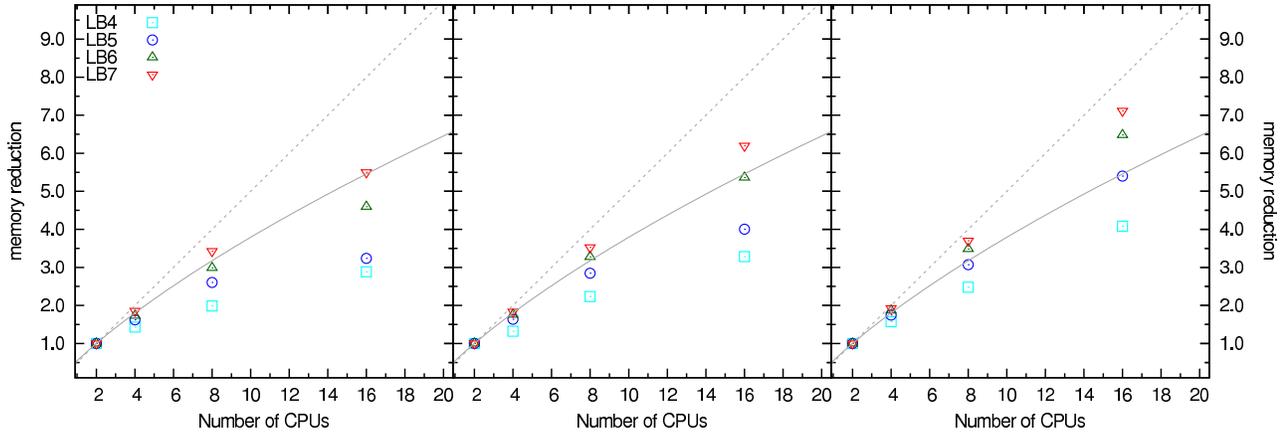}
  \caption{Here we show the achieved memory reduction as defined by
equation~\ref{eq:memredux} of the parallel runs. For guidance,
the curves for linear scaling are shown as a dashed (gray) curve and
additionally, we show the expected ideal memory scaling with a solid
gray line (see \S\ref{subsec:res_lb} for more details).}
  \label{fig:scalability}
\end{figure*}

In figure~\ref{fig:scalability} we closer investigate the achieved
memory saving $R(n)$ with the number of \MPI-processes $n$, defined
as
\begin{equation}
R(n) = M(1)/M(n)
\end{equation}
where $M(n)$ of course refers to the amount of memory required for one
process when using $n$ CPUs.  We correct the memory information for the
contribution of the domain grid (for the choices of parameters it
consumes $64\,\mathrm{MB}$); note that this
correction is not done in figure~\ref{fig:scalability_abs} presented
above.  We chose to do this here to better visualize the actual gain in
the parallel version, which is close to what is expected, when the
relative contribution of the domain grid to the memory usage is small.
That is not quite the case in our test simulations, however this is true
for large simulations.

Note that linear scaling cannot be expected due to the duplicated boundary
zones.  We show this in figure~\ref{fig:scalability} for the ideal case
memory reduction expected including the boundary zones (solid gray line;
for the case of \LB4).  We can estimate this by comparing the amount of
cells each process is assigned to as
\begin{equation}
  R(n, N) = \left(n^{-1/3} + \frac{2}{N} \right)^{-3}
  \label{eq:memredux}
\end{equation}
where $N \equiv 2^{\LB}$ is the number of cells in one dimension on the
\LB-grid and $n$ is the number of CPUs.  Note that for this estimate it
is assumed that the amount of work that needs to be done per cell is
same for all cells and in such this is only an ideal case estimate.  We
can see that this curve is tracked closely in larger boxes when the
distribution of particles is more uniform than in small boxes.

It is important to keep in mind that this holds for simulations with
$256^3$ particles. In fact, the analysis could still be performed in
reasonable time ($\sim 30$ minutes, depending on the box size) and
with modest memory requirements ($\sim 2-3$GB) on a single CPU.  However
a single CPU approach would be unfeasible already for a $512^3$
simulation and completely out of the question for larger simulations.

\subsection{Scalability}
\label{subsec:scalability}

So far we have investigated the scaling behaviour for simulations with a
fixed particle resolution.  It is now interesting to ask how the
algorithm scales when increasing the number of particles in the
simulation.  To this extent we use the simulations of set~2 (see
table~\ref{tab:simulations} for the particulars) which were performed on
the same initial conditions represented with three different particle
resolutions.  All analyses have been performed on
\textsc{hlrbii} at the LRZ M\"unchen.

According to the results derived in \S\ref{sec:testing}, we used the
analysis parameters as $\DomRef = 64$, $\DomRef = 5.0$, $\RefRef = 5.0$
and $\LB = 7$. We used the option to divide the run of \AHF\ into a
splitting and an analysing step (cf. \S\ref{subsubsec:caveats}) and used
$4$ ($16$, $64$) CPUs for the splitting of B936lo (B936me, B936hi).
For the subsequent analysing step, we only changed the number of CPUs for
B936hi from $64$ to $12$.\footnote{Note that this means that a team of
$12$ analysis tasks worked on the $64$ chunks produced in the splitting
step.}  In table~\ref{tab:timing} we present the
full timing information reported by the queueing system for the
different jobs; note that the wall time for B936hi is quite large
compared to the wall time used to B936lo and B936hi,
however this is due to the fact that only $12$ CPUs where used to
process the $64$ chunks generated in the splitting step.

\begin{figure}
  \includegraphics[width=84mm]{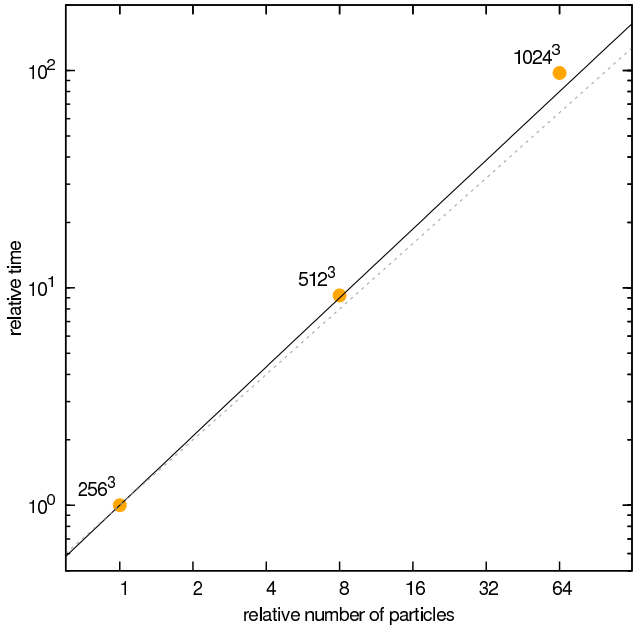}
  \caption{The relative CPU time as reported by the queueing system is
shown depending on the relative simulation size.  The data points are
normalized to the $256^3$ particle representation and we show a linear
scaling with a dashed gray line and an $N\log N$ scaling with a solid
line.  The data points are labeled with the simulation resolution they
correspond to.}
  \label{fig:simsizescaling}
\end{figure}

We also present the scaling in figure~\ref{fig:simsizescaling} were we
show the required CPU time for all three analyses normalized to B936lo.
Note that the expected $N \log N$ scaling is tracked quite remarkably.
It should be noted though that the large box size represents the ideal
situation for \AHF, the scaling will be not as good for smaller box
sizes.

\begin{deluxetable}{lccccc}
  \tablecaption{Timing results for the analyses of set~2. \label{tab:timing}}
  \tablehead{
     \colhead{Name} &
     \colhead{$T^\mathrm{CPU}_\mathrm{an} [h]$} &
     \colhead{$T^\mathrm{Wall}_\mathrm{an} [h]$} & 
     \colhead{$\mathrm{nCPU}_\mathrm{an}$} &
     \colhead{$T^\mathrm{Wall}_\mathrm{sp} [m]$} &
     \colhead{$\mathrm{nCPU}_\mathrm{sp}$}
  }
  \startdata
  B936lo & $ 0.40$ & $0.10$ & $ 4$ & $1.42$ & $ 4$ \\
  B936me & $ 3.74$ & $0.33$ & $16$ & $3.25$ & $16$ \\
  B936hi & $39.42$ & $3.04$ & $12$ & $5.73$ & $64$

  \enddata
  \tablecomments{The subscript `an' refers to quantities related to the
actual analysis, where the subscript `sp' labels quantities related to the
splitting of the simulation volume.  The superscript `CPU' labels the
required CPU time, whereas `Wall' notes the wall clock time.  Note the
times for the analysis process are given in hours, whereas the times for
the splitting are in minutes.}
\end{deluxetable}

\section{Summary \& Conclusions}
\label{sec:conclusions}

We have introduced and described the halo finding tool \AHF, which is
freely provided for use to the community\footnote{Download at
\url{http://www.aip.de/People/AKnebe/AMIGA/}}.  \AHF\ natively
reads \GADGET\ \citep{Springel2001a,Springel2005} format as well
as \AMIGA\ binaries and can deal with large simulation data.
Furthermore the adequate inclusion of  star and gas particles is
supported as well.  \AHF\ requires only very few parameters, in fact,
only five (three) parameters need are to be specified for the parallel
(serial) version.  We have shown in \S\ref{subsec:domgrid} that the size
of the domain grid (\DomGrid) and the refinement criterion on the domain
grid (\DomRef) have only a very marginal impact on the results, reducing
the number of relevant parameters to 3 (1).

However, the refinement criterion on the refined grids (\RefRef) does
influence the completeness of the derived halo catalog.  We discussed
this in detail in \S\ref{subsec:refcrit} finding that to be complete for
halos containing less than $50$ particles, a refinement criterion of
$\lesssim 3.0$ must be chosen.  However, as we show in
figure~\ref{fig:scaling_refref} this increases the memory requirement by
$\sim 40$ per cent and the runtime by a factor of $2-3$.  This will be
mostly interesting for analyses of snapshots at high redshift and low
particle resolution.

Furthermore we have shown in \S\ref{sec:testing} that the number of CPUs
involved in the analysis only has no impact on the derived properties
when the \LB\ parameter (\S\ref{subsec:test_parallel}) is chosen
carefully.  However, given a suitable choice of the \LB\ parameter
a reasonable scaling is achieved (\S\ref{subsec:res_lb}).  This is
especially important for the memory scaling, as this is the key allowing
for the analysis of billion-particle simulations.  In fact he have shown
in \S\ref{subsec:scalability} that, given a good choice of parameters, \AHF\
scales very well with increasing particle resolution.  We have shown
this explicitly for a $1024^3$ simulation in this paper and also have
successfully employed \AHF\ previously on $512^3$ simulations
\citep[e.g.][]{Knollmann2008}.

To summarize the choice of recommended parameters:
\begin{itemize}
  \item We suggest to use a \DomGrid\ of $64$,
  \item a \DomRef\ of $5.0$ and 
  \item a \RefRef\ of $5.0$ to achieve trustworthy results for halos
made up by more than $50$ particles.
  \item To achieve untainted results in the parallel version, the \LB\
level should be chosen in such a way that the relation given in
equation~\ref{eq:boundary} holds.
  \item  The number of CPUs should be chosen as small as possible, the
limiting factor here is the available memory.
\end{itemize}

As we have shown in \S\ref{subsec:res_lb}, the mismatch in runtime
between the separate tasks can become significant.  We provide a way
around this by a two-staged approach: First, the independent volume
chunks including the boundary are constructed on all CPUs.  Then those
will be dumped to disk and can be analysed in serial (or multiple at
once, as the chunks are completely independent).  This approach becomes
important for inhomogeneous matter distributions (small box sizes) and
large number of particles ($>512^3$).  This feature is available in the
public version, for details contact the authors.

\acknowledgements

SRK and AK acknowledge funding through the Emmy Noether Programme of the
DFG (KN 755/1).
AK is supported by the MICINN through the Ramon y Cajal programme.
Simulations have been performed at the Centre for Astrophysics \&
Supercomputing, Swinburne University and the Astrophysical Institute
Potsdam, we thank Chris Power and Christian Wagner for providing the
snapshots to us.  The analyses have been conducted in part on a local machine
at the AIP, whereas most analyses have been performed on the
\textsc{damiana} cluster at the Albert-Einstein Institute in Golm and
the \textsc{hlrbii} at the LRZ M\"unchen  within the context of the
AstroGrid-D project.  SRK would like to thank Steve White for his help
in using the grid infrastructure.
Part of this work was carried out under the HPC-EUROPA project
(RII3-CT-2003-506079), with support of the European Community - Research
Infrastructure Action under the FP6 ``Structuring the European Research
Area'' Programme.

\appendix
\section{AHF's unbinding procedure}
\label{app:AHFunbinding}

In order to remove gravitationally unbound particles we have to obtain
the (local) escape velocity $v_\mathrm{esc}(r)$ at each particle
position. As $v_\mathrm{esc}$ is directly related to the (local) value
of the potential, $v_\mathrm{esc} = \sqrt{2\left| \phi \right|}$,
we integrate Poisson's equation (under the assumption of spherical
symmetry):
\begin{equation}
 \Delta \phi(r) = \frac{1}{r} \frac{\dif}{\dif r}
                  \left(
                     r^2 \frac{\dif\phi} {\dif r} 
                  \right)
                 = 4 \pi G \rho (r)
\end{equation}
The first integral reads as follows
\begin{equation}
  r^2\frac{\dif\phi}{\dif r} 
  - \left[
       r^2\frac{\dif\phi}{\dif r}
    \right]_{r=0} 
= 4 \pi G \int_0^r \rho(r') r'^2 \dif r' 
= G M(<r)
\end{equation}
This equation shows that $\dif\phi/\dif r \propto M(<r)/r^2$ and
hence $ r^2 \dif\phi/\dif r \rightarrow 0$ for $r \rightarrow 0$. We are   
therefore left with the following first-order differential equation
for $\phi(r)$:
\begin{equation}
 \frac{\dif\phi}{\dif r} = G \frac{M(<r)}{r^2}
\end{equation}
Another integration leaves us with
\begin{equation} \label{eq:AHFphi}
 \phi(r) = G \int_0^r \frac{M(<r')}{r'^2} \dif r' + \phi_{0}
\end{equation}
This time we need to calculate $\phi_0$. We do this
by requiring $\phi(\infty)=0$:
\begin{eqnarray}
 \phi(\infty) & = & G \int_0^\infty \frac{M(<r')}{r'^2} \dif r' + \phi_0 \\
              & = & G \int_0^{r_\mathrm{vir}} \frac{M(<r')}{r'^2} \dif r' +
                    G \int_{r_\mathrm{vir}}^{\infty} \frac{M(<r')}{r'^2}
                    \dif r' + \phi_0 \\
              & = & G \int_0^{r_\mathrm{vir}} \frac{M(<r')}{r'^2} \dif r' +
                    G M_\mathrm{vir} \int_{r_\mathrm{vir}}^\infty
                    \frac{1}{r'^2} \dif r' + \phi_0 \\
              & = & G \int_0^{r_\mathrm{vir}} \frac{M(<r')}{r'^2} \dif r' +
                    G \frac{ M_\mathrm{vir} }{r_\mathrm{vir}} + \phi_0
\end{eqnarray}
and hence
\begin{equation}
 \phi_0  =  - \left( G \int_0^{r_\mathrm{vir}} \frac{M(<r')}{r'^2} \dif r' +
G \frac{ M_\mathrm{vir} }{r_\mathrm{vir}} \right)
\end{equation}
Note that we assume that the halo is truncated at
$r_\mathrm{vir}$ when evaluating the integral $\int_{r_\mathrm{vir}}^{\infty}
\frac{M(<r')}{r'^2} \dif r'$.  An initial guess for particles belonging
to a (sub-)halo in a cosmological simulation is based upon the adaptive
mesh hierarchy of the simulation code \AMIGA, as we have described in
\S\ref{subsec:ahf}.  Unbound particles are removed iteratively where we
integrate equation~\ref{eq:AHFphi} along a list of radially ordered
particles; the same holds for obtaining $\phi_0$ that has to be
re-evaluated for every new iteration.

\bibliographystyle{apj}
\bibliography{ahf}

\end{document}